# Cost-effective implementation of the Paris Agreement using flexible greenhouse gas metrics

(Short title: Flexible emission metrics for the Paris Agreement)

Katsumasa Tanaka[1,2,3,*], Olivier Boucher[1], Philippe Ciais[2], Daniel J. A. Johansson[4], Johannes Morfeldt[4]

[1] Institut Pierre-Simon Laplace (IPSL), Centre national de la recherche scientifique (CNRS)/Sorbonne Université, Paris, France

[2] Laboratoire des Sciences du Climat et de l'Environnement (LSCE), IPSL, Commissariat à l'énergie atomique et aux énergies alternatives (CEA)/CNRS/Université de Versailles Saint-Quentin-en-Yvelines (UVSQ), Université Paris-Saclay, Gif-sur-Yvette, France

[3] Center for Global Environmental Research, National Institute for Environmental Studies (NIES), Tsukuba, Japan

[4] Division of Physical Resource Theory, Department of Energy and Environment, Chalmers University of Technology, Gothenburg, Sweden

* Corresponding author

Email: katsumasa.tanaka@lsce.ipsl.fr

**Abstract**

Greenhouse gas (GHG) metrics, that is, conversion factors to evaluate the emissions of non-$CO_2$ climate forcers on a common scale with $CO_2$, serve crucial functions upon the implementation of the Paris Agreement. While different metrics have been proposed, their economic cost-effectiveness has not been investigated under a range of pathways, including those temporarily missing or significantly overshooting the temperature targets of the Paris Agreement. Here we show that cost-effective metrics for methane that minimize the overall cost of climate mitigation are time-dependent, primarily determined by the pathway, and strongly influenced by temperature overshoot. The Paris Agreement will implement the conventional 100-year Global Warming Potential (GWP100), a good approximation of cost-effective metrics for the coming decades. In the longer term, however, we suggest that parties consider adapting the choice of common metrics to the future pathway as it unfolds, as part of the global stocktake, if cost-effectiveness is a key consideration.

## INTRODUCTION

Aligning climate policies with the goals of the Paris Agreement implies to revisit the concept of GHG metrics. Emission metrics offer a simple way to quantify the combined climate impacts from the emissions of a mix of radiatively active gases and aerosols, without requiring a model. Specifically, a metric serves as an exchange index to convert the emission of a non-$CO_2$ climate forcer to a so-called "$CO_2$-equivalent emission" for climate policy purposes (Fig. 1). Emission metrics have, however, been a subject of debate and evaluation (*1, 2*), since the inception of GWP (*3, 4*) in 1990. GWP100, the most widely used metric today, equates the emissions of different climate forcers with respect to the radiative forcing integrated over 100 years after a pulse emission (*2*). The consistency of GWP100 with policy goals has been questioned from physical and economic perspectives, and many alternatives have been proposed (*5-15*). The choice of metrics also reflects the priority for issues of concern (*16, 17*), influencing particularly how much the emissions of $CH_4$, a short-lived climate forcer (SLCF), can be reduced relative to those of $CO_2$, which is of considerable importance in high $CH_4$-emitting countries (Supplementary Fig. 1). While the scientific debate on metrics continues (*18, 19*), GWP100 has been adopted as the common metrics for the Paris Agreement implementation (*20*). Adopting common metrics for all parties to the Paris Agreement was a vital advance in the development of the so-called Paris Rulebook since it will enable cross-party comparisons of progress toward respective nationally determined contributions (NDCs), allow assessing the effect of specific mitigation actions and policies put in place, and permit trading of credits between parties. However, it is unclear how costly this decision will be if GWP100 is the metric that will be permanently used for future mitigation strategies.

Here we evaluate the use of GWP100 and other emission metrics in climate mitigation strategies on the basis of their cost-effectiveness, one of the key principles of the United Nations Framework Convention on Climate Change (UNFCCC) (Article 3 of ref. (*21*)), as well as a guiding principle for climate mitigation pathways in the previous Intergovernmental Panel on Climate Change (IPCC) Assessment Reports. We use a simple globally aggregated Integrated Assessment model (IAM) (*22, 23*) that accounts for distinct atmospheric characteristics and emission abatement costs of $CO_2$ and $CH_4$ among other features. This model is used to derive a cost-effective pathway for a given temperature target and explore how the use of a metric can influence the pathway and mitigation costs. The analysis was performed for a contrasted range of pathways toward the Paris Agreement temperature target levels, including those overshooting the target levels temporarily. Small overshoots (i.e. exceedance warming up to 0.1 or 0.2 °C) have been considered in previous studies (*24-27*); however, larger overshoots – reflecting the current trend (i.e. 3 °C warming by 2100 (*28, 29*)) – have not been considered before despite their increasing likelihood.

Examining metrics in light of overshoot possibilities illuminates the path-dependency of metric cost-effectiveness. According to our IAM, the theoretically most cost-effective metric for $CH_4$ (i.e. the ratio of the shadow prices of $CO_2$ and $CH_4$), which is known to be time-dependent under stabilization pathways (*7*), shows larger temporal variations under overshoot pathways. The path- and time-dependency of cost-effective metrics led us to explore further how the choice of conventional $CH_4$ metrics such as GWP100, GWP50, and GWP20 can be adapted in policy setting to evolving future pathways. We show that the *adaptive* use of metrics – changing from GWP100 to other shorter time horizon GWPs in the future – can save mitigation costs, compared to the *permanent* use of GWP100. Nevertheless, GWP100 appears to be a reasonably good approximation of the most cost-effective metrics for the next few decades. Hence, our study suggests that the UNFCCC and parties to the Paris Agreement should consider periodically assessing the appropriateness of the choice of GHG metrics as future options unfold to pursue optimal implementation of emissions abatement options. Such assessments could be considered as part of the recurring global stocktake processes within the UNFCCC, where future mitigation actions and ambition levels are discussed among parties, since they are related to potential revisions of the adopted GHG metrics.

Aligning emission metrics with emerging mitigation strategies

The framing of climate policy-relevant research has changed drastically with the adoption of the Paris Agreement in 2015 (*30*), which calls for holding the global warming well below 2 °C, pursuing efforts to limit the warming to 1.5 °C relative to preindustrial levels (Article 2.1), and achieving global net zero anthropogenic GHG emissions during the latter half of this century (Article 4.1). The temperature targets have led to massive research efforts on low temperature stabilization pathways, as assessed in the IPCC Special Report "Global warming of 1.5 °C" (SR15) (*29*). The SR15 highlighted the need for global net zero anthropogenic $CO_2$ emissions by 2050 to achieve the 1.5 °C goal. An increasing number of countries, municipalities, sectors, and individual firms adopted so-called "carbon neutrality", or zero emissions, for mid-century.

In contrast to these ambitious goals, there is a bitter reality: the current NDCs imply that the global warming may exceed 3 °C by the end of this century (*28, 29*). While there are pressures to ratchet up the NDCs (*31*), even planned mitigation efforts may face challenges upon implementation. The demonstrations by the Yellow Vests ("gilets jaunes" in French), triggered in November 2018 by the opposition to rising fuel tax in France, were illustrative of showing the difficulties of implementing fair climate policies. In the first half of 2020, the global $CO_2$ emissions were significantly dropped relative to the corresponding 2019 level as a result of the pandemic of coronavirus disease 2019 (COVID-19) (*32*). To enable emission reductions for a longer term, green investment is promoted as part of the recovery packages in response to the economic downturn cause by the pandemic (*33*), but it is yet unclear how the long-term emission trend will unfold. Given these circumstances, as argued by others before (*34-36*), it is imperative to consider a broad range of pathways, not only *stabilization* (or non-overshoot) pathways toward the 2 and 1.5 °C targets but also *overshoot* pathways, under which the temperature exceeds the target levels significantly before eventually settling there. Pathways overshooting the 1.5 °C target and staying below 2 °C can be interpreted as being in line with the Paris Agreement's long-term goal, while pathways overshooting the 2 °C target should be seen as being in violation of the goal. Nevertheless, such overshoot pathways are considered plausible if ambition levels are not rapidly increased in the next decades. GHG metrics were evaluated for various temperature and forcing targets (*18, 24, 26, 37*), with a few studies analyzing metrics directly applied to the net zero GHG target (*23, 38*). But, to our knowledge, metrics have not been investigated under a range of plausible overshoot pathways in relation to the Paris Agreement goals.

At the twenty-fourth session of the Conference of the Parties (COP24) of the UNFCCC, GWP100 was adopted as the common metrics to be used for the implementation of the transparency framework of the Paris Agreement (paragraph 37 of the Annex to Decision 18/CMA.1 (*20*)). The transparency framework includes guidelines for parties' GHG inventories, scenario analyses, and reporting on implemented policies and measures, with the purpose of tracking progress with the implementation of each party's NDC and informing the global stocktake (paragraph 1 of the Annex to Decision 18/CMA.1 (*20*)). To be more specific, the guideline states that the GWP100 values shall be used as provided by the IPCC Fifth Assessment Report (AR5) (or a subsequent IPCC report upon future agreement), an update from those provided by the IPCC Second or Fourth Assessment Report (SAR or AR4) currently in use at the UNFCCC (*39*). The COP24 decision gave an option for using other metrics in addition to GWP100, noting the Global Temperature change Potential (GTP) (*8*) as an example, which equates the emissions of different climate forcers with respect to the final *temperature change* at the *end* of a chosen time horizon after a pulse emission.

While the decision has been made on the common metrics for the implementation of the Paris Agreement, it is important to emphasize that the COP24 decision contains a provision for reviewing and updating the guidelines for the transparency framework by 2028 at the latest (paragraph 3 of Decision 18/CMA.1 (*20*)) that could allow a shift from GWP100 towards another type of common metrics, provided that parties wish to pursue such an update that goes beyond the current decision. It should be further noted that the topic of common metrics is still listed as one of the sub-items of Methodological issues in the draft provisional agenda for the fifty-second session of the Subsidiary Body of Scientific and Technological Advice (SBSTA52) (*40*), an advisory body to the UNFCCC on scientific and technical matters. The negotiations at SBSTA did not reach conclusions at the previous session and the agenda item is automatically pushed to the next session planned for 2021. Nevertheless, the draft text (*41*) proposed by the Chair of the previous session suggests that parties recognize the importance of synchronizing metrics used for different purposes within the UNFCCC, guided by the decision made for the transparency framework of the Paris Agreement. At the same time, the draft text shows openness to considering future scientific findings on metrics in IPCC AR6, as well as views of parties and observers on the implications of the choice of metrics for climate change policy. In response, a group of scientists made a voluntary submission to UNFCCC to encourage SBSTA52 to initiate and conclude a review of common metrics, via a dialogue between policymakers and the research community (*42*).

## RESULTS

Cost-effective metrics under stabilization and overshoot pathways

We present a first analysis of the implications of using different emission metrics under stabilization and overshoot pathways. To start with, we show the Global Cost-effective Potential (GCP) (*7, 43*), the most economically efficient metric that can be derived from a cost-effective pathway of interest, which serves as a benchmark in the analysis that follows. Unlike GWP and GTP, which require only a physical model to be estimated, GCP requires calculations of both climate change and mitigation costs with an IAM. Our study employs the Aggregated Carbon Cycle, Atmospheric Chemistry, and Climate model (ACC2) version 4.3 (*22, 23*) (Methods), a simple IAM that describes major physical and biogeochemical processes in the global earth system, as well as the economic relationships between mitigation levels of GHG emissions and associated costs. More specifically, ACC2 consists of a global climate and carbon cycle model of reduced complexity combined with an economic module containing marginal abatement cost (MAC) curves for $CO_2$, $CH_4$, and $N_2O$ (*37, 44*) (Supplementary Fig. 2). This model allows us to generate cost-effective emission pathways for given temperature targets under a single modeling framework, unlike other more complex IAMs that are not dynamically coupled with a climate and carbon cycle model (*45*). The model describes primarily global aspects, providing no details in regional and sectoral changes. The temporal resolution of the model is one year. Limitations associated with the MAC curve approach are discussed in Methods and elsewhere (*46, 47*).

We generated 2 °C stabilization as well as 2 °C and 1.5 °C overshoot pathways on the basis of a cost-effectiveness principle by minimizing the net present value costs of abating the GHG emissions over time as calculated from the MAC curves (Fig. 2a; Supplementary Table 1; Supplementary Fig. 3). Fig. 2a shows five such illustrative pathways in comparison to those considered in SR15 (*48*) (Methods). We assume two different overshoot levels requiring the warming not to exceed the respective target after 2100 or 2150. For example, we applied a 2 °C target beginning in 2100 in the cost-effective pathway calculation to derive our 2 °C pathway with overshoot until 2100. Thus, the length and magnitude of temperature overshoot were not externally set in our analysis but rather an outcome of our internal cost-effective pathway calculations. The time scale beyond 2100 is essential for investigating a broad range of overshoot pathways aligning the current trend toward a 3 °C warming before the end of this century. We interpret the two temperature targets stated in Article 2 as stabilization targets (i.e. stable temperatures) in our analysis. Note that no 1.5 °C stabilization pathway is included in our analysis because the 1.5 °C target cannot be achieved without overshoot in our model unless we relax the abatement constraints (Methods) (*23*). The earliest possible year to achieve the 1.5 °C target is 2084 after overshoot under our default model assumptions. Sensitivities of these pathways to underlying assumptions are discussed where appropriate below.

Under each of these five pathways, we calculated GCP, a metric most consistent with the cost-effectiveness principle. GCP is defined as the ratio of the willingness to pay for emitting an additional unit of a gas of interest to that of $CO_2$ at each point in time under a cost-effective pathway, reflecting the ratio of the emission shadow prices between the two gases. We implicitly assume a globally connected emission market for $CO_2$, $CH_4$, and $N_2O$. In this analysis, we focus on the outcome of $CH_4$, a potent GHG whose atmospheric response time is substantially shorter than that of $CO_2$ (*49*). The outcome of $N_2O$, a long-lived GHG, is different but does not strongly influence the overall outcome, so it will not be analyzed here.

Our calculations show that, under the 2 °C stabilization pathway, the $CH_4$ GCP rises over time until the temperature reaches 2 °C (Fig. 2b) (*7, 12, 37*). GCP for $CH_4$ is larger than GTP100 throughout the period and becomes larger than GWP100 (i.e. 28 in the IPCC AR5) after 2040. After the stabilization in 2063, GCP stays at an approximately constant level near GWP50. For the overshoot pathways, the rise in the $CH_4$ GCPs occurs later and more drastically than under the stabilization pathway. GCPs grow after the temperature peak, until immediately before the temperature returns to the target levels, with a peak exceeding GWP20. The rise in the $CH_4$ GCPs is associated with the priority given to $CH_4$ mitigation to lower the temperature to the target level (Supplementary Fig. 4). This is because of the rapid effect $CH_4$ mitigation has on the temperature, but other factors also come into play. Such deep $CH_4$ abatement prior to the stabilization was not observed under the stabilization pathway. Once the target is met, the temperature does not have to be reduced further in our cost-effectiveness approach, which no longer requires as much $CH_4$ mitigation as before, resulting in the abrupt drop in GCPs. The temperature is slightly decreased further after the stabilization due to the inertia of the physical earth system before it finally settles at the target level.

Our results from the five illustrative pathways of Fig. 2 show that the $CH_4$ GCP is time-dependent, rises till stabilization occurs – particularly strongly so under the overshoot pathways – and then becomes stable after stabilization. The long-term evolution of GCP depends on the type of pathways. Overshoot pathways imply larger changes in GCP. These results also suggest that the evolution of GCP is largely determined by the year when the stabilization is eventually achieved (related insight in ref. (*12*)) and relatively insensitive to the target level (Supplementary Fig. 5), but not at all influenced by the year of the temperature peak, which is sometimes proposed for the end of the time horizon for a dynamic GTP (*8*). We further explore the sensitivity of GCP with respect to the assumptions on the equilibrium climate sensitivity (2 °C and 4.5 °C, with 3 °C by default), the discount rate (2% and 6%, with 4% by default), and the MAC curves (one case assuming a 50% higher $CO_2$ MAC curve and 50% lower $CH_4$ and $N_2O$ MAC curves than the respective standard MAC curves and the opposite case assuming a 50% lower $CO_2$ MAC curve and 50% higher $CH_4$ and $N_2O$ MAC curves than the respective standard MAC curves) (Methods). GCP is sensitive to the assumptions on the discount rate and the equilibrium climate sensitivity, while being less sensitive to the MAC curves (*12*); however, the behavior of GCP described above generally holds under different assumptions in our IAM (Supplementary Fig. 6).

<u>The cost-effectiveness of GWPs and GTPs</u>

The time-dependent GCP is the most cost-effective metric by construction, as it is derived directly from the cost-effective pathway calculation. We now estimate the economic implications of continuing with the time-invariant GWP100 currently in use and agreed to be used in the Paris Agreement implementation. The economic costs of using GWP100 can be calculated by imposing the metric in the pathway calculation, namely, on the ratios of the marginal abatement costs of associated gases at each point in time. That is, one keeps the $CH_4$ marginal abatement cost in each year to be larger than the $CO_2$ marginal abatement cost in the same year by a factor of the $CH_4$ GWP100. Likewise, one applies the $N_2O$ GWP100 to fix the ratio of the $CO_2$ and $N_2O$ marginal abatement costs over time (but this constraint on $N_2O$ has only small impact on the overall results). Thus, the use of metrics poses additional constraints in the pathway calculation, giving rise to higher mitigation costs than those without the use of metrics. The cost increment as a result of the metric use, relative to the lowest costs without the use of metrics (or equivalently with the use of GCP), is analyzed as the "cost of metrics" in our study. Note that the use of metrics influences the emission and temperature pathways and can even affect the feasibility of the temperature target. Thus, we made further assumptions to analyze the cost of metrics as explained in Methods. The methodology described above follows several previous studies (*26, 50-52*), but there are variations in the methodologies for metric cost calculations, requiring attention when the outcomes are compared. For example, some previous studies (*24, 25*), using more complex IAMs than the one employed here, used $CO_2$-equivalent emission targets derived from different metrics, instead of directly constraining the ratios of relevant gas prices, to estimate the costs of using metrics.

Previous studies showed that the use of GWP100 does lead to some but not a significant disadvantage in terms of global total abatement costs under stabilization pathways (including small overshoot pathways) (*24-27, 50-52*). While we confirm this finding (Fig. 3; 1.4% [0.6%, 2.5%] higher global total mitigation costs (till 2200) than the least cost case, with sensitivity ranges in square brackets with respect to the climate sensitivity, the discount rate, and the MAC curves (Methods)), we find that the additional costs of using GWP100 become larger under overshoot than stabilization pathways (e.g. 4.0% [2.5%, 5.5%] under the 1.5 °C large overshoot pathway). The costs of using GWP20 are also higher under overshoot than stabilization pathways (e.g. 10.9% [9.0%, 11.7%] under the 1.5 °C large overshoot pathway). It is worth noting that the use of GWP20, which is sometimes promoted due to concern over the near-term warming (*53*), yields larger long-term total costs than that of GWP100 under all pathways, including sensitivity cases (Supplementary Fig. 7), reflecting high $CH_4$ abatement driven by this metric. The use of other time-invariant metrics, in particular GTP100, creates additional costs of over 10% in all cases (Supplementary Fig. 8, except for those using the discount rate of 2%). For longer time horizons, the use of GWP500 and GTP500 leads to additional costs of over 8 and 20%, respectively, under all pathways (Supplementary Table 2; Supplementary Figs. 9 and 10). The high costs associated with the use of long time horizon metrics including GTP100 are due to the need for more $CO_2$ abatement at a higher cost to compensate for $CH_4$ emissions valued too low (*26, 27*).

We further show the "cost-effective" time horizon for GWP or GTP, which reduces the total mitigation costs as much as possible if implemented throughout the period, a variant of ref. (*10*). We repeated the metric cost calculations by changing the metric time horizon for $CH_4$ and $N_2O$ between one year and 500 years with a one-year interval (a five-year interval above 150 years). We then identified the cost-effective time horizon leading to lowest costs under each pathway. The GWP and GTP values for $CH_4$ and $N_2O$ with a time horizon from one to 500 year(s) were calculated based on Section 8.SM.11 of the IPCC AR5 (for metrics *without* inclusion of climate-carbon feedbacks for non-$CO_2$). The calculated metric values (Supplementary Fig. 11) reproduced those reported in Tables 8.A.1 and 8.SM.17 of the IPCC AR5. The values of GWPs and GTPs with long time horizons need to be taken cautiously because the uncertainty in metric values increases with time horizon (*49, 54*).

The cost-effective time horizon depends on the pathways, ranging between 58 [45, 86] and 110 [92, 118] years for GWP and between 29 [25, 34] and 38 [36, 40] years for GTP (Fig. 3). In both metric cases, the longer the period before the stabilization year occurs, the longer the cost-effective time horizon is. The range is more confined for GTP than for GWP, largely reflecting the different correspondence between time horizons and metric values (Supplementary Fig. 11). There is however still about 4% additional costs even with the use of the cost-effective time horizon under the overshoot pathways (for both GWP and GTP). This finding suggests that, no matter which time horizon is chosen, the use of a single GWP or GTP departs significantly from cost-effectiveness under overshoot pathways. This arises from the fact that the temporal variations in GCPs under overshoot pathways cannot be well approximated by a single static GWP or GTP. It may be worthwhile to consider the dynamic GTP whose time horizon is kept until the time of stabilization (*24, 26*) because it captures the rising trend toward the point of stabilization, which can contribute to cost saving. It is known that a dynamic GTP can exceed GCP values under stabilization pathways (*12, 37*) and may follow the sharp rise in GCP under overshoot pathways. However, we did not analyze it here because it is unclear how to apply a dynamic GTP in the post-stabilization period.

The additional costs discussed above may appear rather modest and, indeed, the choice of pathways has a much larger impact on the absolute costs than the choice of metrics (Supplementary Fig. 9c, 10c). Nevertheless, the choice of metrics strongly influences the cost distributions over time and across gases (with the exception that the impacts on the $N_2O$ abatement costs, emissions, and concentrations are generally insignificant (Supplementary Figs. 12 to 16)). Previous studies (*24, 25, 27*) reported significant regional and sectoral impacts from the choice of metrics, despite relatively small global impacts. Our results imply these regional and sectoral variations in cost would be even more significant under overshoot pathways.

## Best available metrics among the IPCC set

From a practical perspective, it is useful to interpret our theoretical cost-effective outcome through the eyes of well-established metrics from the IPCC AR5. We thus translated the GCP results (Fig. 2b) in terms of GWPs (or GTPs) with representative time horizons of 20, 50, and 100 years. In other words, we selected "best available GWPs (or GTPs)" from AR5, whose values are most proximate to our GCPs in absolute terms at each point in time, under the assumption that these metric values from the IPCC will not change in the future. In the experiment, we treated GWPs and GTPs separately because these are structurally different.

We mainly analyze the results from GWPs, which are predominantly used in policies.

Under the 2 °C stabilization pathway, best available metrics change from GWP100 to GWP50 (or from GTP50 to GTP20) shortly before 2050 (Fig. 4). Under the overshoot pathways, changes in best available metrics are more drastic than under the stabilization pathway, reflecting the larger changes in GCPs with overshoot. Differences in the transitions of best available metrics come from different stabilization years. Another interesting outcome is that GWP100 is chosen from the onset under all pathways, a robust finding except regarding the assumed set of available metrics (Methods; Supplementary Figs. 17 to 20). If we assume six metrics to choose from (i.e. GWP500, GWP200, GWP100, GWP50, GWP20, and GWP10), it chooses GWP500 or GWP200 at the beginning, although these long time horizons are rarely used in practice. On the other hand, if we assume just two metrics in the basket of choice (i.e. GWP100 and GWP20), it chooses GWP100 for most of the period under all pathways, except for some periods shortly before the temperature stabilization (Supplementary Fig. 17).

How cost-effective is it to move from a fixed approach permanently using GWP100 to a flexible approach allowing future revisions of the metric choice? While the additional costs of *continuously* changing metrics, like the dynamic GTP, have been investigated previously (*24, 26*), those of *discretely* changing metrics, which is arguably more policy relevant, have not been considered before. We show here, for the first time, that it is more cost-effective to shift from the fixed to the flexible approach under all pathways, taking also account of their sensitivity cases (Fig. 5; all outcomes, that is, five representative outcomes shown in large circles, as well as sensitivity outcomes expressed as error bars in horizontal and vertical directions, are consistently on the right of the 1:1 line). The cost improvement is larger under overshoot than stabilization pathways. The assumed set of available metrics influence such benefit. Under the 1.5 °C large overshoot pathway (as the most extreme case), choosing from a set of six GWPs instead of the default set of three GWPs reduces the additional costs from 2.3% down to 0.46%. On the other hand, choosing from a set of two GWPs increases the costs to 2.5% (but still below 4.0%, the costs of using only GWP100 under this pathway). The assumption on the discount rate has a larger impact on the results than those on the climate sensitivity and the MAC curves – however, the economic benefit for the flexible approach, relative to the permanent use of GWP100, is retained in all cases.

We further find that the flexible approach using best available metrics is more cost-effective than *any* fixed approach, including one using the metric with the cost-effective time horizon. In the former case, the choice is limited to three GWPs but it can be changed at any time. In the latter case, the choice can be any GWP with a time horizon of 1 to 500 year(s) but it cannot be changed with time. Under all pathways including sensitivity cases (a total of 28 cases), the flexible use of best available GWPs is less costly than the fixed use of a single GWP with the cost-effective time horizon (Fig. 3; Supplementary Fig. 7). The same finding was obtained from a corresponding experiment using GTPs, except for one case assuming the discount rate of 2% (Fig. 3; Supplementary Fig. 8). Our results point to a fundamental limit associated with the fixed use of metrics and further support the flexible use of metrics to ensure cost-effectiveness under a range of mitigation pathways.

## DISCUSSION

### Implications for the Paris Agreement implementation

Our new findings support the provision of flexibility in the implementation of the Paris Agreement, especially towards metrics with shorter time horizons that can reflect future changes in long-term pathways. We argue that this aspect of mitigation planning could be considered at the recurring global stocktake processes (first in 2023 and then every five years) by including an assessment of the cost-effectiveness of GHG metrics. Following the COP24 decision (Decision 19/CMA.1 (*20*)), the global stocktake serves as a tool for increasing the ambition needed from parties to achieve the goals of the Paris Agreement. The process is supported by a technical assessment to include opportunities for enhanced mitigation action (two sessions in November 2022 and July 2023 for the first global stocktake occurrence; the deadline for materials to be considered in the technical assessment expected to be August 2022). Promoting cost-effective emission abatement is important in this process since it can identify such opportunities and may enable increased ambition levels from parties. Hence, we suggest that an assessment of the cost-effectiveness of GHG metrics for future mitigation strategies could be included as an input to the technical assessment supporting the global stocktake. When deemed necessary, the global stocktake could recommend a revision of the Paris Agreement's transparency framework to shift towards GWPs with shorter time horizons.

In fact, a change in metric values is not unprecedented at the UNFCCC. A small revision is already implicit in the COP24 decision to update GWP100 values to those in AR5 (and those in any future IPCC reports), which means that the $CH_4$ GWP100 will be revised from 21 (SAR) or from 25 (AR4) to 28 (AR5) (or to 34 (AR5), depending on the treatment of climate-carbon feedbacks (*55*)). But the long-term metric revisions indicated by our analysis are of a different nature and much larger than the proposed update of GWP100 values. This further highlights the need for GWP estimates for different time horizons to be continuously provided by the IPCC, irrespective of scientific advances in other types of metrics.

It is important to note that our suggestion is not in conflict with the use of GWP100 in the coming decades. Rather, we argue for a long-term benefit for allowing flexibility in the future choice of metrics, fundamentally because the future pathway is unknown due to the inherent uncertainties in the climate system and the unpredictability of future social, economic, and technological developments; technically because a stabilization year, which is crucial for determining the metric, is not explicitly given in the Paris Agreement text. Even if a 2 and 1.5 °C pathway is followed within this century, and assuming that global cost-effectiveness is an important criterion for guiding international climate policy, the metric would still be time-dependent in response to changing mitigation priorities, particularly under overshoot pathways.

We further note on the newly proposed metrics such as GWP* (*15*), or its variation (*56*), which can enhance the accuracy of how $CO_2$ and SLCF emissions are aggregated in terms of the implied warming (*18, 57*). The GWP* approach in essence compares $CO_2$ emissions of a certain year with the rate of SLCF emissions over a preceding period (e.g. for 20 years). While such new metrics are intensively debated in recent literature, however, it is yet unclear to what extent they would require revisions of the Paris Agreement itself to accommodate the need for considering SLCF emissions occurring in preceding decades, if they are applied to the mitigation context that our analysis deals with. Shifting among conventional GWPs with different time horizons seems to be a more practical way of maintaining its integrity while ensuring a cost-effective implementation of the Paris Agreement.

Our approach assumes a global actor, without considering the potentially heterogeneous behaviors of individual actors toward time-dependent metrics in the absence of perfect knowledge (*58*). Nevertheless, cost-effective metrics presented here serve as a useful benchmark, against which more transparent metrics like GWPs can be evaluated. Our cost-effective metrics and the corresponding shifts of GWPs depend on assumptions in the IAM required to calculate future pathways. However, as presented through various sensitivity analyses, the cost advantage for adapting the metric to evolving future was shown to be robust under a broad range of pathways. Our findings suggest that, while the possibility of metric revisions should not impede the political progress toward the Paris Agreement implementation nor disturb market-based mechanisms relying on metrics, there should be room for consideration of metric revisions in the future potentially as part of the global stocktake processes within the UNFCCC. The flexible use of metrics is more inclusive than the permanent use of GWP100, better serving for a broad range of pathways under changing circumstances, including non-ideal overshoot cases.

## METHODS

Model ACC2

ACC2 represents four domains of the global earth system: i) physical climate system, ii) carbon cycle system, iii) atmospheric chemistry system, and iv) economy system. The first three domains are described in the next paragraph and the last one in the paragraph that follows. We keep the model description succinct here, only describing the aspects most pertinent to our present analysis. The current model was developed from earlier simple climate models (*59, 60*) and produces an equivalent output with the one used in ref. (*23*). The model is written by the General Algebraic Modeling System (GAMS) language.

The physical climate system is represented by an energy balance model coupled with a heat diffusion model (*22, 61*). Radiative forcing agents considered in the model include $CO_2$, $CH_4$, $N_2O$, $SF_6$, 29 species of halocarbons, tropospheric and stratospheric $O_3$, and stratospheric water vapor. Aerosol forcing is separated by three terms: the direct effect of sulfate aerosols, the direct effect of black carbon and organic aerosols, and the indirect effects of all aerosols. The $CH_4$ lifetime is influenced by OH, $NO_x$, CO, and VOC. Note that each forcing term is calculated separately without any gas aggregation using metrics like GWP100. The global carbon cycle is provided by a box model: four boxes for the coupled atmosphere-ocean and another four for the land. Saturation of ocean $CO_2$ uptake under rising atmospheric $CO_2$ concentrations is modeled through the thermodynamic equilibrium of carbonate species in the ocean. The $CO_2$ fertilization of the land biosphere is parameterized by a commonly used β factor. No climate-carbon feedbacks are assumed in our analysis; that is, carbon cycle processes are assumed to be insensitive to the temperature change. The equilibrium climate sensitivity is fixed at 3 °C, within the 1.5 – 4.5 °C range suggested by the IPCC AR5 (in the Thematic Focus Elements 6). Other uncertain parameters such as those related to aerosol forcing and $CO_2$ fertilization are optimized based on a Bayesian approach using historical observations such as global-mean temperature changes and atmospheric $CO_2$ concentrations (*62*). The interdependencies among the parameter estimates are considered, including the one between the climate sensitivity and the aerosol forcing strength (*63*). The optimization is performed by using the CONOPT3, a nonlinear optimization solver provided with GAMS.

The economy module is used to estimate the costs of mitigating $CO_2$ (fossil fuel origin), $CH_4$, and $N_2O$ emissions based on a first-order method using global MAC curves (*37, 44*) (Supplementary Fig. 2). The MAC curves are assumed time-invariant and given as a function of the abatement level (in percent) of the respective gas relative to an assumed baseline level (i.e. the International Institute for Applied Systems Analysis (IIASA) Greenhouse Gas Initiative (GGI) A2r baseline scenario (*64*)). The fossil fuel $CO_2$ MAC curve is based on the output of the GET energy system model (*44*), which was simulated iteratively under different future trajectories of the carbon price. Although the carbon price for a given level of emission reductions should be time-dependent, we use a mathematical function to approximate the data collected from 2060 to 2100 and apply it as the MAC function throughout the period in our analysis. Limitations associated with the fixed MAC curve approach are partially but imperfectly mitigated by the constraints on the temporal changes in the abatement level to account for the technological change and socioeconomic inertia associated with emission abatement. Namely, the rate of change in the abatement level (i.e. first derivative) is kept below 4% per year for all three gases, implying a limit for the technological change; furthermore, the rate of abatement change (i.e. second derivative) is below 0.4% per year, mimicking socioeconomic inertia. These first- and second-derivative constraints limit the extent of the MAC curves that can be used in the near term. The implication of these constraints is that, if the abatement starts in 2020, the abatement level can reach up to 20% in 2030, 60% in 2040, and 100% in 2050 (i.e. zero emissions from fossil fuel $CO_2$). These constraints allow larger changes in the abatement level than those found in the 450 ppm $CO_2$eq scenarios in the AR5 database (Supplementary Figs. 12 to 14 of ref (*23*)). The maximum abatement levels for $CO_2$ (fossil fuel origin), $CH_4$, and $N_2O$ are assumed at 112%, 70%, and 50%, respectively. The abatement potential for $CO_2$ can exceed 100% primarily because the IAM, on which our $CO_2$ MAC curve is based, considers bioenergy combined with carbon capture and storage as an option in the mitigation portfolio (*44*). Our approach is kept simple and works under the assumption that $CO_2$ and non-$CO_2$ mitigation measures are interchangeable, which is partially true given the necessity to finance mitigation actions but may also break down for measures involving co-reduction of GHGs. Our MAC curve approach does not capture GHG abatement measures entailing net negative costs that have however not been implemented due to non-economic factors. The emissions of all other gases and pollutants including $CO_2$ from land use change are prescribed without cost calculations (i.e. GGI A2r 480 ppm $CO_2$eq stabilization). The discount rate is assumed at 4% by default. We analyze the pathway until 2200, going beyond the 2100 time frame commonly analyzed. The long time frame is required to capture overshoot pathways under which temperatures will not return to 2 °C or lower during this century.

Pathways and metric costs

The stabilization and overshoot pathways were derived from the cost-effective calculation method described above. We let the model to determine the abatement levels of three gases over time under the abatement constraints (i.e. first- and second-derivative constraints as well as the upper limits of abatement levels) to arrive at a pathway that meets a policy objective while minimizing the global total costs of mitigation. Two temperature target levels (2 °C and 1.5 °C) and three temperature pathway profiles (non-overshoot, medium overshoot (till 2100), and large overshoot (till 2150)) were considered. Temperature overshoot emerges as a consequence of how the temperature target is implemented: the target is assumed effective only after a certain point in time in the future.

It should be noted that the reference pathways against which the additional costs of using metrics were calculated are not identical to the illustrative pathways in Fig. 2. All metric cost calculations in our study did not use the abatement constraints, that is, the first- and second-derivative constraints as well as the upper limits of abatement levels, which were used to derive the illustrative pathways. These constraints influence the metric cost calculations and in some cases make the pathway infeasible because they can be too restrictive when applied together with metrics. In order to keep consistency, the abatement constraints were also not used in the reference pathways to derive the cost of metrics. However, the overall pathways and mitigation costs without the use of abatement constraints are not substantially different from those with the use of such constraints, except for cases with large overshoot. Another exception is the periods when the mitigation starts and when the target is met (Supplementary Table 2; Supplementary Figs. 12 to 16). In such periods, particularly under the overshoot pathways, abatement levels can change drastically in the absence of the abatement constraints, requiring careful interpretation.

Sensitivity analysis

We consider the following three sources of uncertainty: equilibrium climate sensitivity, discount rate, and MAC curves. The equilibrium climate sensitivity is assumed to be 3 °C by default, with sensitivity cases of 2 °C and 4.5 °C. In comparison to the 1.5 – 4.5 °C range, the uncertainty in equilibrium climate sensitivity indicated by the IPCC AR5, our analysis did not consider climate sensitivity below 2 °C as suggested by a previous study using ACC2 (*62*) and recent lines of evidences (*65*). The discount rate is set at 4% by default and assumed at 2% and 6% in sensitivity cases, spanning a typical range considered in cost-effectiveness analyses (i.e. 5 to 6% in line with market interest rates), as well as low discount rates suggested by recent literature (*66*). The uncertainty in MAC curves is generally large and a related study reports an uncertainty range of ±50% in the MAC curves (*52*). We consider two cases changing the priority of $CO_2$ and non-$CO_2$ mitigation alternately: one case assuming a 50% higher $CO_2$ MAC curve and 50% lower $CH_4$ and $N_2O$ MAC curves and the opposite case assuming a 50% lower $CO_2$ MAC curve and 50% higher $CH_4$ and $N_2O$ MAC curves. Note that we vary the assumption on the $N_2O$ MAC curve for consistency, but this has little influence on the overall results. In the sensitivity analysis, we vary the assumptions on these uncertainties just one by one from the default and do not vary more than one assumption at a time due to the computational burden, yielding a total of seven cases including the default case. A larger number of sources of uncertainty were considered in the historical inversion of the physical part of the model. But in the metric cost analysis demanding more computational resource, we focus on the equilibrium climate sensitivity, as the most important uncertain parameter in the physical earth system, while acknowledging that other parameters, including those related to climate-carbon cycle feedbacks, can also be important. Note that, with the climate sensitivity of 4.5 °C, the 2 °C stabilization pathway and the 1.5 °C medium overshoot pathway are not feasible with the abatement constraints put in place. As a result, these two pathways are not considered also in the metric cost analysis when the climate sensitivity is set at 4.5 °C, even though these pathways are feasible without the abatement constraints. The largest climate sensitivity that makes these target pathways feasible are 3.4 °C in both cases.

In the analysis of best available metrics, we further consider the sensitivity of the assumed set of available metrics. The default set of time horizons considered for GWP and GTP are 100, 50, and 20 years. As a comparison, the IPCC AR5 lists the values of GWP20, GWP100, GTP20, GTP50, and GTP100 for a number of climate forcers in Table 8.A.1. AR5 also reports GWPs and GTPs with a time horizon of 100, 50, 20, and 10 years for a limited number of climate forcers in Table 8.SM.17. In contrast, the previous IPCC Assessment Reports up to AR4 present the values of GWPs with a time horizon of 500, 100, 20 years (GTP values are only in AR5). We consider the following alternative sets of available time horizons for GWP and GTP: two time horizons (100 and 20 years) and six time horizons (500, 200, 100, 50, 20, and 10 years). In the sensitivity analysis, we choose a time horizon from two or six available time horizons, the $CH_4$ GWP (or GTP) of which is closest to the GCP at each point in time. The chosen time horizon is also be used for $N_2O$ GWP (or GTP) as done in the default analysis. In this exercise, we refer to the AR5 metric values *without* inclusion of climate-carbon feedbacks for non-$CO_2$ gases. Note that same metric values are assumed in the future period in our analysis. The IPCC metric values have in fact changed over the assessment cycles due to several compounding and competing factors, including improvement in scientific understanding and changing background atmospheric conditions (Section 8.7.2.1 of the IPCC AR5). The metric values will probably be revised also in the future, but such changes are impossible to predict.

Scenarios obtained from the IPCC SR15

Data for the temperature pathways considered by SR15 were downloaded from the Integrated Assessment Modeling Consortium (IAMC) 1.5 °C Scenario Explorer hosted by IIASA (*48*). The following five classes of pathways are considered in SR15: Below-1.5°C, 1.5°C-low-OS, 1.5°C-high-OS, Lower-2°C, and Higher-2°C (Table 2.1 of the IPCC SR15). Out of 222 scenarios in the five classes, 92 temperature pathways are available from the Scenario Explorer website for download (accessed on 8 February 2019). Most of them are given for the period 2005–2100 with five or ten-year intervals. There are two temperature pathways indicating temperatures significantly higher than 2°C during this century (one peaking at 2.3 °C in 2080 and the other at 2.66 °C in 2090), which were removed from our analysis. Temperature data in SR15 use the 1850-1900 mean temperature as a reference. The 1850-1900 mean temperature in our model is 0.0294 °C based on the inverse calculation (*22, 62*). Thus, the SR15 temperature pathways shown in Fig. 2 are adjusted accordingly to account for the difference in the base temperature.

**SUPPLEMENTARY MATERIALS**

Supplementary Materials are available for this paper.

**Acknowledgments**

The authors are grateful for comments and suggestions from Seita Emori, Jan Fuglestvedt, Brian O'Neill, Hideo Shiogama, Kayo Tanaka, Junichi Tsutsui, and Tokuta Yokohata, which were useful for this study.

**Funding**

This work benefited from State assistance managed by the National Research Agency in France under the "Programme d'Investissements d'Avenir" under the reference "ANR-19-MPGA-0008". K.T. was supported by a Make Our Planet Great Again (MOPGA) Short-Stay grant in France (N° dossier: 927201A). O.B. acknowledges support from the Ministère de la transition écologique et solidaire (MTES) of France through the "Convention relative à l'attribution d'un appui financier au bénéfice des services climatiques."

**Author contributions**

K.T. and O.B. conceived this study. K.T. led the study. K.T., O.B., P.C. and D.J.A.J. designed the experiment. K.T. performed the analysis. K.T., O.B., P.C., D.J.A.J., and J.M. analyzed the results. K.T. drafted the manuscript, with contributions from all co-authors.

**Competing interests**

The authors declare no competing financial interests.

**Data and materials availability**

Correspondence and requests for materials should be addressed to K.T.

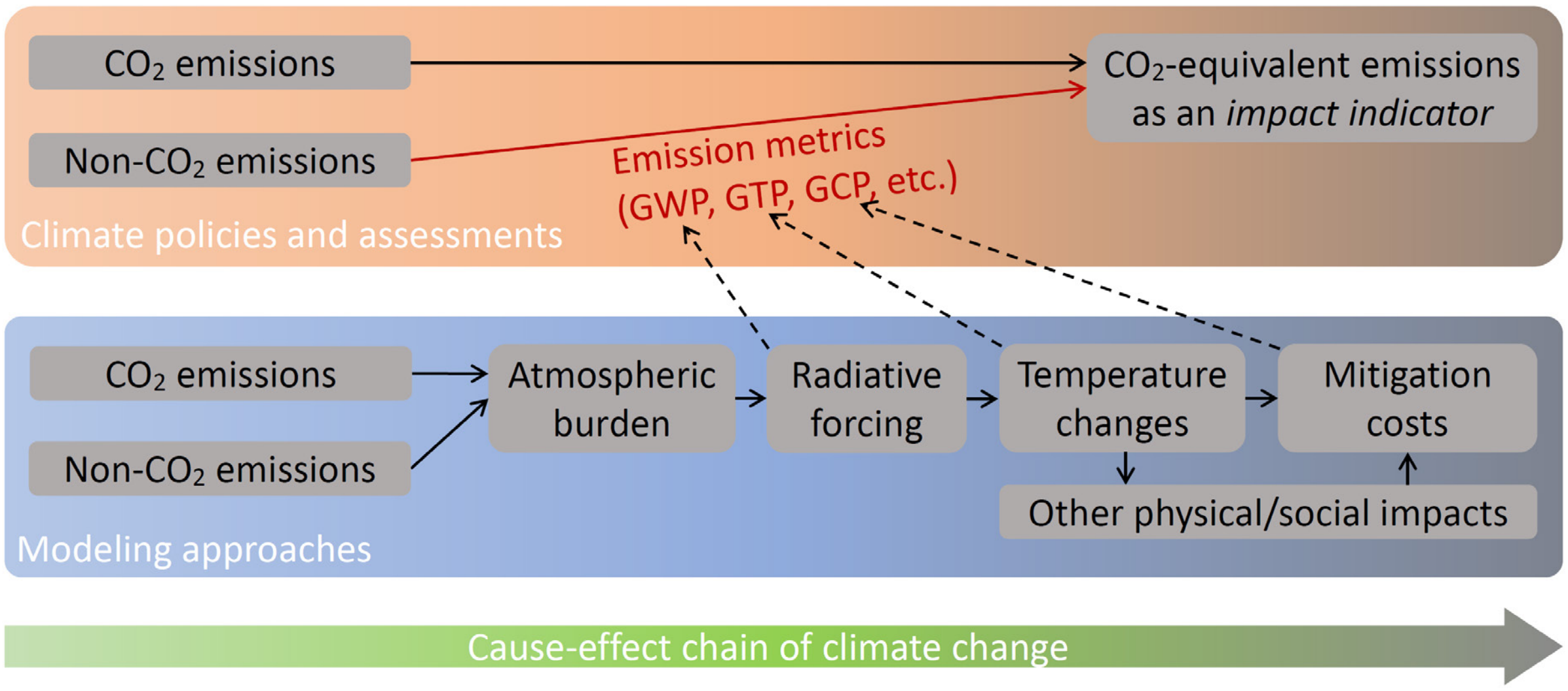

**Fig. 1. The use of greenhouse gas emission metrics in climate policies and assessments and its relations to modeling approaches.** The figure shows our interpretation that climate policies and assessments (orange-to-grey area) regard $CO_2$-equivalent emissions as a surrogate indicator of climate impacts, while modeling approaches (blue-to-grey area) look more directly into the temperature change and other physical and social impacts as an indicator of climate impacts. Grey boxes show factors, such as emissions and temperature change, along the cause-effect chain of climate change from left to right (green-to-grey bar), following Fig. 8.27 of the IPCC AR5. Solid arrows represent cause-effect relationships between such factors. The arrow where emission metrics are applied is highlighted in red. If GWP is used, the conversion from non-$CO_2$ emissions to $CO_2$-equivalent emissions implicitly uses radiative forcing calculations using models (dashed arrow). Likewise, if GTP and GCP are used, the $CO_2$-equivalent conversion relies on temperature and mitigation cost calculations using models, respectively (dashed arrows). For the purpose of clarity, only first-order relationships are shown. Temporal and spatial aspects are suppressed in the figure.

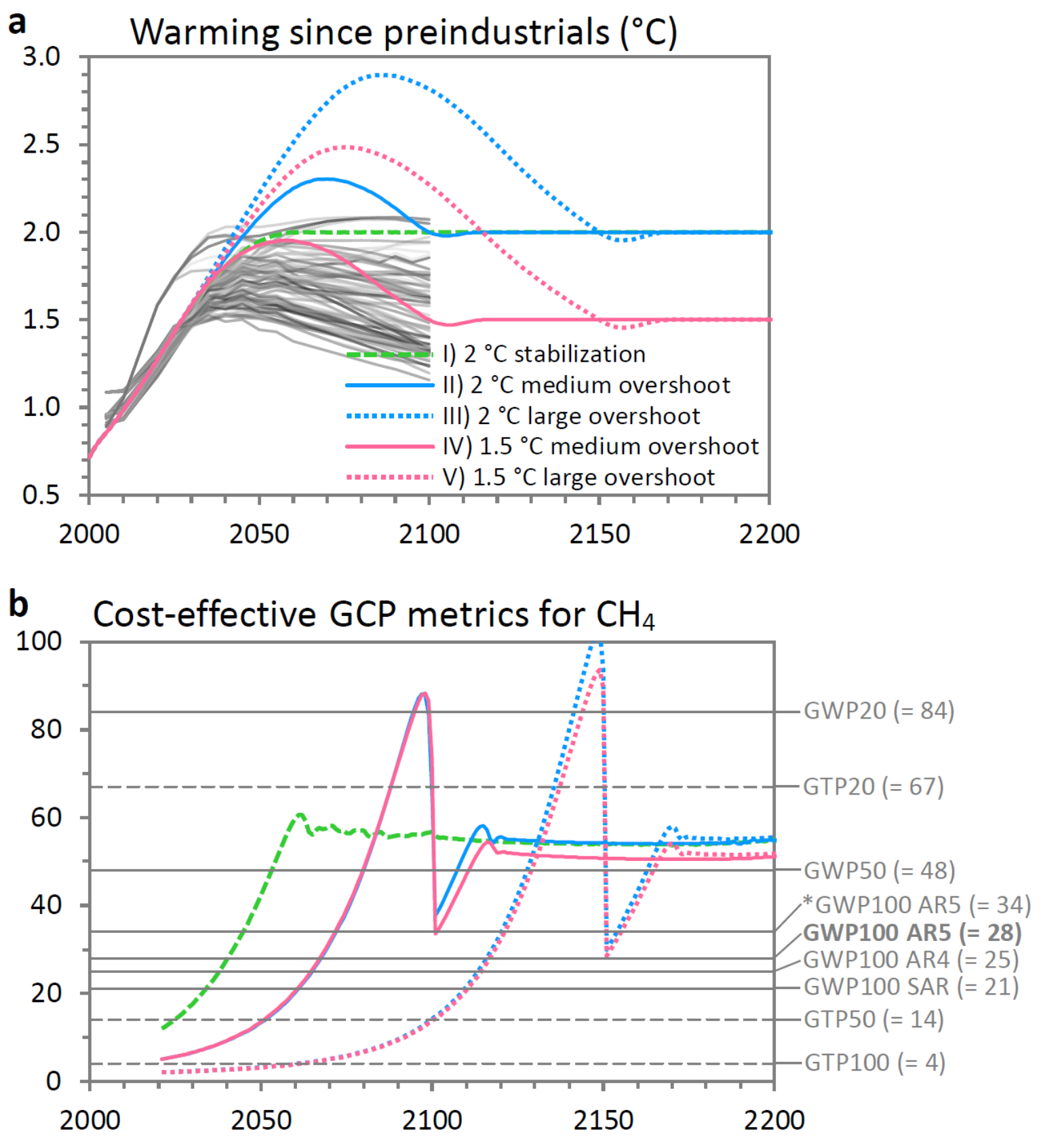

**Fig. 2. Temperature stabilization and overshoot pathways and cost-effective GCP metrics for $CH_4$.** In panel **a**, dashed green lines show the case in which the 2 °C target is achieved without overshoot. Blue and pink lines indicate the cases in which the temperature stabilizes at the 2 and 1.5 °C warming levels after overshoot, respectively. Solid and dotted lines correspond to cases where overshoot is assumed unavoidable before 2100 and 2150, respectively (termed "medium" or "large" overshoot, respectively, given "small" overshoot in the IPCC SR15). Grey lines in the background indicate the range of temperature pathways considered in SR15 (Methods). In panel **b**, the $CH_4$ GCPs under the five pathways (line designations as in panel **a**) are shown. GWPs and GTPs with the time horizons of 20, 50, and 100 years are shown in solid and dashed grey lines, respectively, as a reference for comparison. The metric values indicated in parentheses are taken from Tables 8.A.1 and 8.SM.17 of the IPCC AR5 (i.e. those *without* inclusion of climate-carbon feedbacks for non-$CO_2$), unless noted otherwise. For GWP100, the panel shows four different values including those from earlier IPCC Assessment Reports. The GWP100 value in AR5 *with* inclusion of climate-carbon feedbacks for non-$CO_2$ is indicated by * and taken from Table 8.7 of AR5.

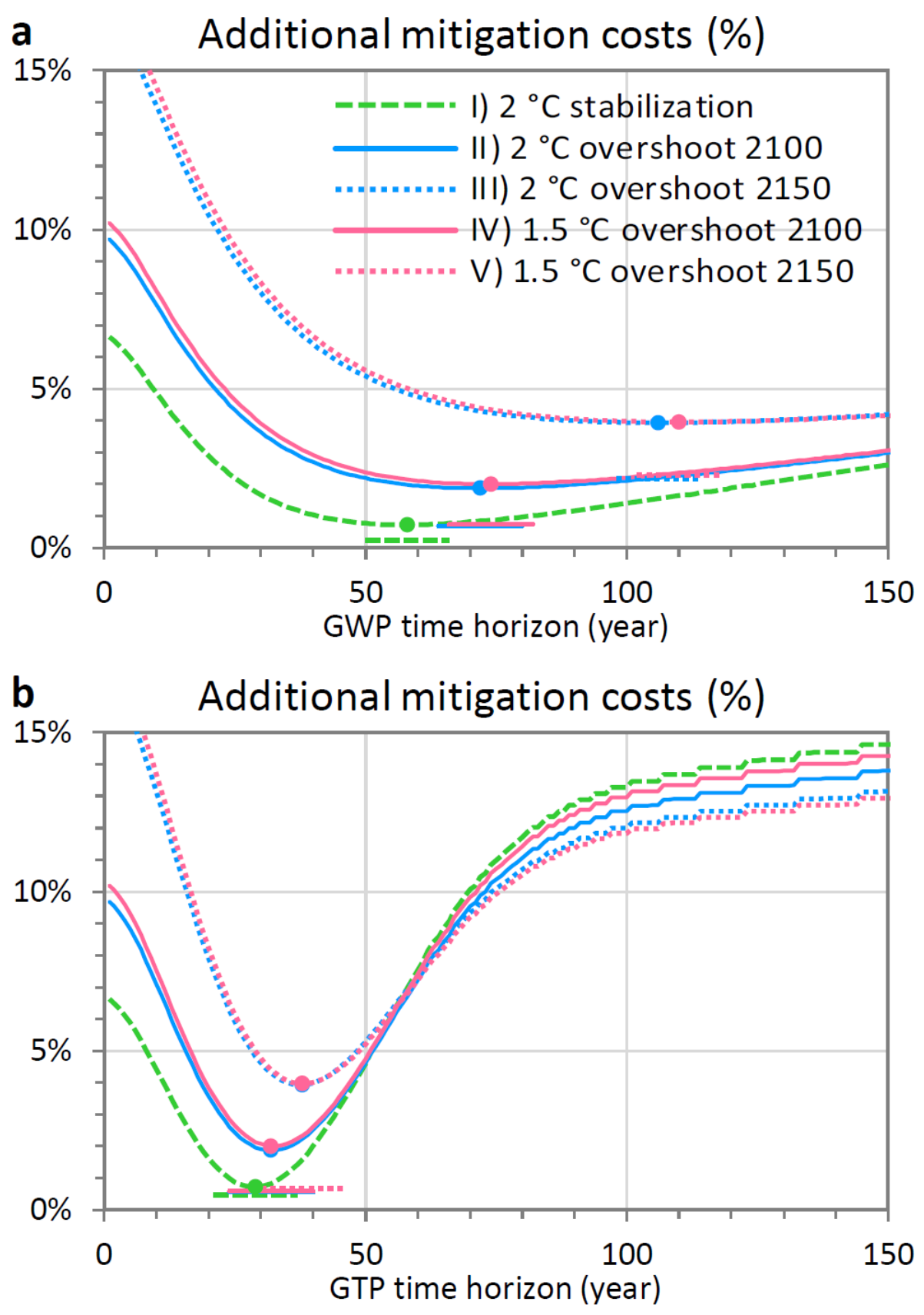

**Fig. 3. Additional mitigation costs of using GWP and GTP with a range of time horizons under the stabilization and overshoot pathways.** The additional mitigation costs (in percent) with the use of GWP and GTP relative to the lowest costs without the use of metrics (or equivalently with the use of GCP) are shown in panels **a** and **b**, respectively. The results with the time horizons between one and 150 year(s) are presented. The minimum under each pathway, which is marked by a filled circle, indicates the cost-effective time horizon and the residual additional mitigation costs. The short horizontal bars vertically aligned with each minimum point indicates the additional mitigation costs of using best available GWPs (from the default set of three metrics GWP100, GWP50, and GWP20) (panel **a**) or best available GTPs (from the default set of GTP100, GTP50, and GTP20) (panel **b**) under each pathway. All horizontal bars follow the legend in panel **a**.

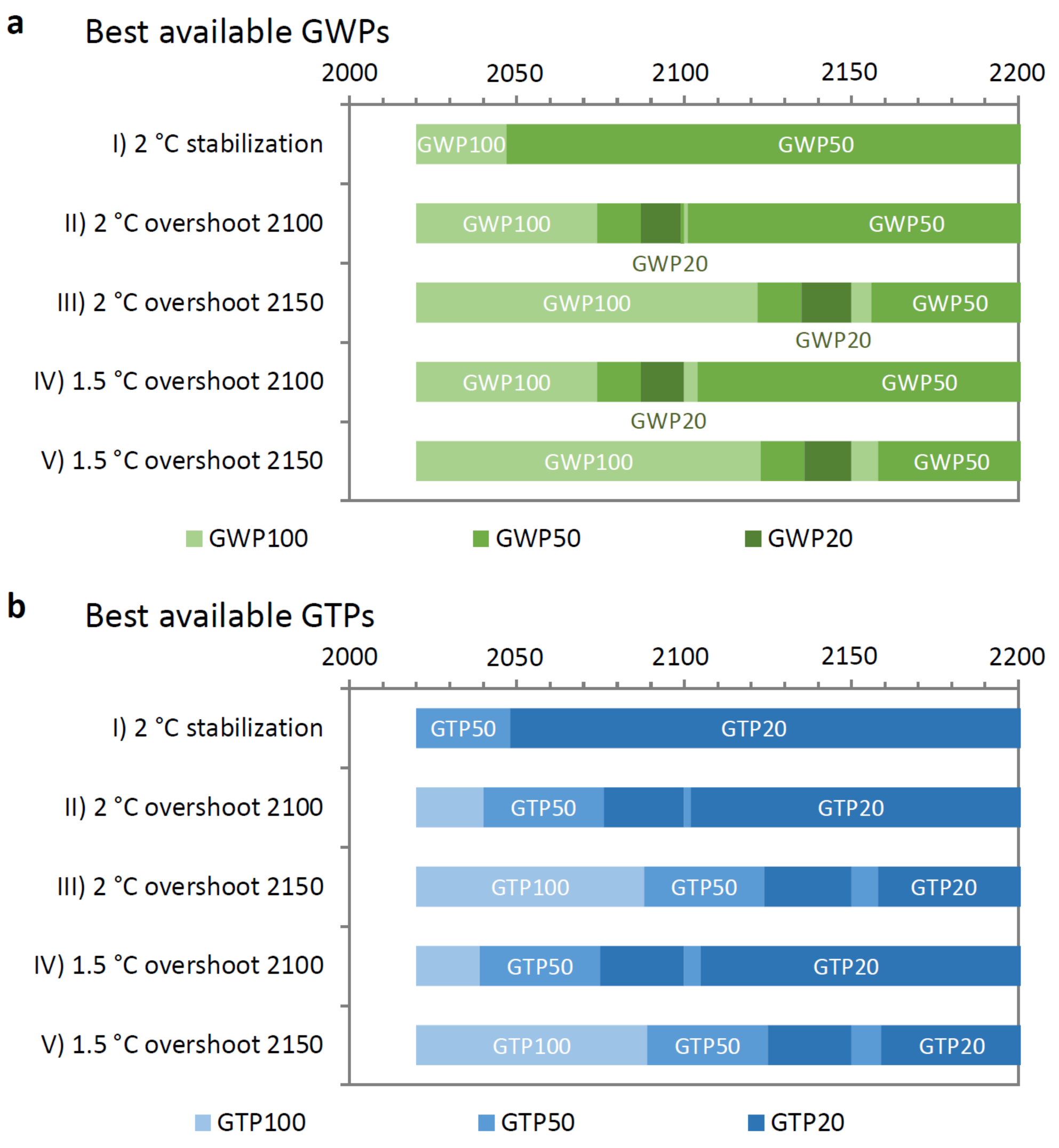

**Fig. 4. Choices of representative GWPs and GTPs most proximate to cost-effective GCP metrics under the stabilization and overshoot pathways.** GWPs and GTPs with three representative time horizons (i.e. 20, 50, and 100 years) from the IPCC AR5 are considered. One of the three GWPs and GTPs (in panels **a** and **b**, respectively), whose value is closest to the corresponding GCP in absolute terms, is shown under each pathway. The color is designated according to the time horizon as indicated in the legend at the bottom of each panel. On the basis of ref. (*67*), we refer to the IPCC metric values *without* inclusion of climate-carbon feedbacks for non-$CO_2$ gases (Tables 8.A.1 and 8.SM.17 of the IPCC AR5), while noting that it is unclear whether the COP24 decision (paragraph 37 of the Annex to Decision 18/CMA.1 (*20*)) refers to metric values with or without inclusion of climate-carbon feedbacks for non-$CO_2$ gases. It should also be noted that the IPCC AR5 does not endorse any metrics assessed.

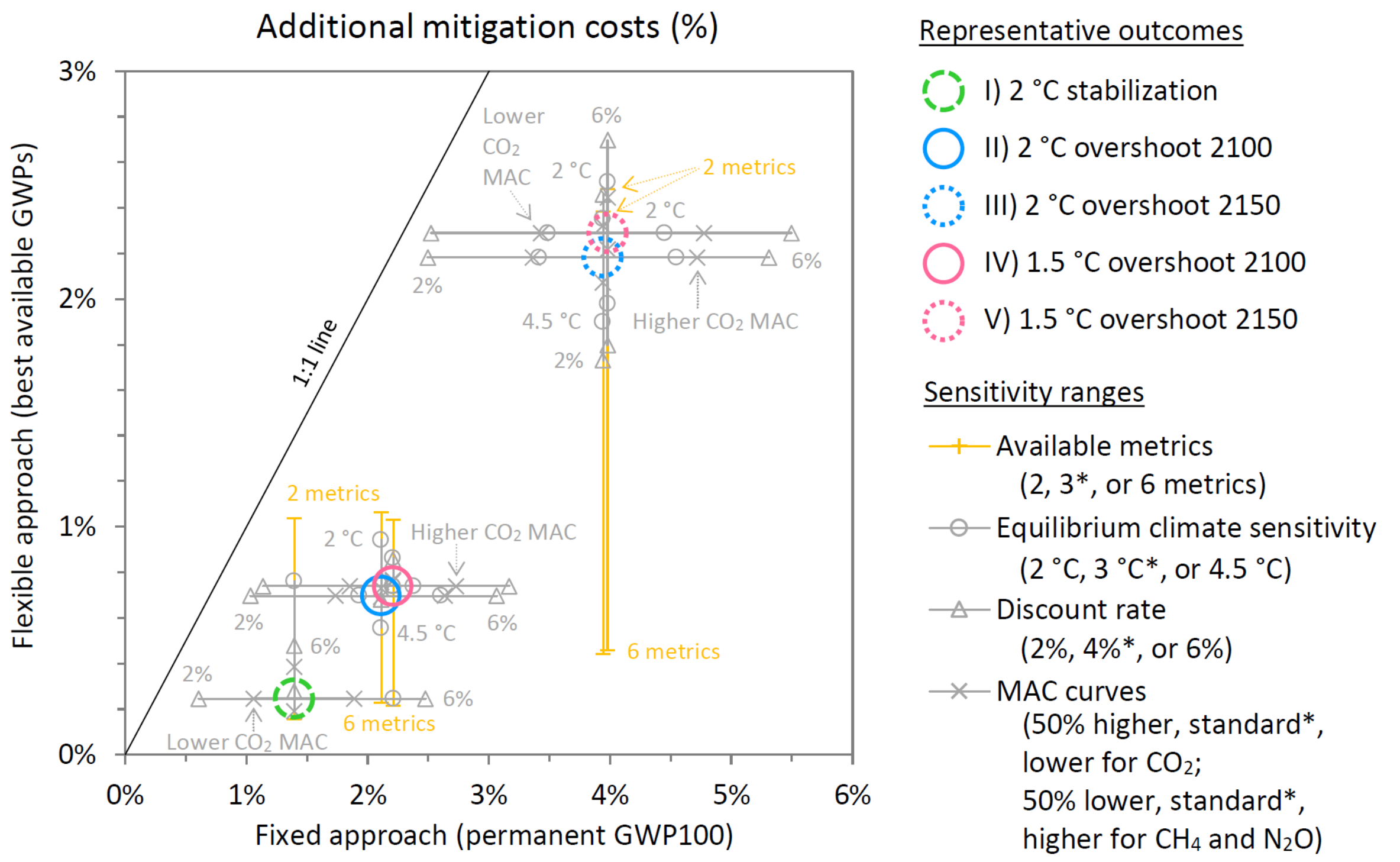

**Fig. 5. Additional mitigation costs of shifting from the permanent use of GWP100 to the more flexible use of GWPs.** This scatter plot shows the additional mitigation costs (in percent) with the permanent use of GWP100 (x-axis) versus those with the use of best available GWPs (y-axis), both relative to the lowest costs without the use of metrics (or equivalently with the use of GCP). The 1:1 line is indicated in black. The outcomes under default assumptions are indicated in large circles as representative outcomes. The sensitivity ranges are shown in both horizontal and vertical directions and characterized by categories. Note that horizontal error bars indicate their respective sensitivity ranges for the fixed approach only. The same goes for vertical error bars indicating sensitivity ranges for the flexible approach only. By definition, sensitivity ranges with respect to the number of available metrics (in yellow) appear only to the vertical direction. In the legend, * indicates the assumption in default. The figure shows that the costs are larger in the cases with fewer available GWPs, low climate sensitivity, high discount rate, or high $CO_2$ MAC curve (and low $CH_4$ and $N_2O$ MAC curves) than those in the respective opposite cases (due to the space limit, labels for sensitivity cases are given only for selected plots). Note that the 2 °C stabilization pathway and the 1.5 °C medium overshoot pathway are not considered in the case of 4.5 °C climate sensitivity since these pathways are infeasible with the high climate sensitivity.

**Cost-effective implementation of the Paris Agreement using flexible greenhouse gas metrics**

**– Supplementary Materials –**

(Short title: Flexible emission metrics for the Paris Agreement)

Katsumasa Tanaka[1,2,3,*], Olivier Boucher[1], Philippe Ciais[2], Daniel J. A. Johansson[4], Johannes Morfeldt[4]

[1] Institut Pierre-Simon Laplace (IPSL), Centre national de la recherche scientifique (CNRS)/Sorbonne Université, Paris, France

[2] Laboratoire des Sciences du Climat et de l'Environnement (LSCE), IPSL, Commissariat à l'énergie atomique et aux énergies alternatives (CEA)/CNRS/Université de Versailles Saint-Quentin-en-Yvelines (UVSQ), Université Paris-Saclay, Gif-sur-Yvette, France

[3] Center for Global Environmental Research, National Institute for Environmental Studies (NIES), Tsukuba, Japan

[4] Division of Physical Resource Theory, Department of Energy and Environment, Chalmers University of Technology, Gothenburg, Sweden

* Corresponding author

Email: katsumasa.tanaka@lsce.ipsl.fr

**Contents**

| | Temperature target | | Outcomes | | | | |
|---|---|---|---|---|---|---|---|
| Cases | *Level* | *Pathway* | *Peak warming (°C) and year* | *Years at or above 1.5 °C* | *Years at or above 2 °C* | *2015-2100 $CO_2$ emissions ($GtCO_2$)* | *2015-2150 $CO_2$ emissions ($GtCO_2$)* |
| I | 2 °C | Stabilization | 2.00 °C (2063) | 2028- | 2063- | 1,469 | 1,601 |
| II | 2 °C | Overshoot before 2100 | 2.30 °C (2070) | 2028- | 2047-2099 | 1,553 | 1,670 |
| III | 2 °C | Overshoot before 2150 | 2.90 °C (2086) | 2028- | 2043-2149 | 2,901 | 2,028 |
| IV | 1.5 °C | Overshoot before 2100 | 1.95 °C (2058) | 2028-2099 | — | 549 | 440 |
| V | 1.5 °C | Overshoot before 2150 | 2.48 °C (2075) | 2028-2149 | 2045-2115 | 1,759 | 836 |

**Supplementary Table 1. Temperature targets and key outcomes of the five illustrative pathways analyzed in this study.** The table shows the temperature target and the pathway feature applied for each pathway and the key outcomes such as the timings and levels of peak warming, the periods of temperature overshoot, and the carbon budgets calculated directly from the model. The carbon budgets presented above account for only $CO_2$ (i.e. non-$CO_2$ components not included) and indicate net emissions.

| a. 2 °C stabilization | Peak Profile | | Overshoot Profile | | Carbon Budget | | Cost distribution (period) | | | | Cost distribution (gas) | | | Total Costs | |
|---|---|---|---|---|---|---|---|---|---|---|---|---|---|---|---|
| Cases | *Peak Warming* | *Peak year* | *Size* | *Duration (year)* | *2015-2100 ($GtCO_2$)* | *2015-2150 ($GtCO_2$)* | *2021-2050* | *2051-2100* | *2101-2150* | *2151-2200* | *$CO_2$* | *$CH_4$* | *$N_2O$* | *Relative* | *Absolute (trillion US$)* |
| i) Reference pathway (without constraints) | 2.00 °C | 2065 | | | 1,476 | 1,608 | | | | | | | | | 75.1 |
| ii) Illustrative pathway (with constraints) | 2.00 °C | 2063 | | | 1,469 | 1,601 | 2.3% | 2.7% | 0.0% | 0.0% | 2.4% | 2.9% | -8.9% | 2.3% | 76.8 |
| iii) GWP100 | 2.00 °C | 2066 | | | 1,363 | 1,453 | 2.9% | 1.8% | -2.5% | -3.4% | 5.3% | -32.9% | -15.1% | 1.4% | 76.1 |
| iv) GWP20 | 2.00 °C | 2067 | | | 1,568 | 1,719 | -1.7% | 4.0% | 4.2% | 4.6% | -5.1% | 82.2% | -23.9% | 2.9% | 77.3 |
| v) GTP100 | 2.00 °C | 2065 | | | 1,038 | 1,027 | 33.6% | 10.3% | -1.8% | -4.7% | 24.6% | -89.3% | -11.4% | 13.3% | 85.1 |
| vi) GTP20 | 2.00 °C | 2067 | | | 1,527 | 1,665 | -2.0% | 2.7% | 1.9% | 2.1% | -3.1% | 49.2% | -18.3% | 1.6% | 76.3 |
| vii) GWP58 (cost-effective time horizon) | 2.00 °C | 2066 | | | 1,442 | 1,555 | -0.6% | 1.5% | -1.1% | -1.5% | 1.2% | -0.9% | -15.1% | 0.7% | 75.6 |
| viii) GTP30 (cost-effective time horizon) | 2.00 °C | 2066 | | | 1,442 | 1,556 | -0.5% | 1.5% | -1.1% | -1.5% | 1.1% | -1.4% | -12.1% | 0.7% | 75.6 |
| ix) Best available GWPs (time-dependent) | 2.00 °C | 2066 | | | 1,451 | 1,576 | -0.3% | 0.6% | -0.8% | -0.9% | 0.9% | -3.9% | -16.1% | 0.2% | 75.3 |
| x) Best available GTPs (time-dependent) | 2.00 °C | 2064 | | | 1,496 | 1,648 | 1.6% | 0.0% | 1.1% | 1.8% | -0.7% | 13.9% | -14.7% | 0.5% | 75.4 |

| b. 2 °C overshoot 2100 | Peak Profile | | Overshoot Profile | | Carbon Budget | | Cost distribution (period) | | | | Cost distribution (gas) | | | Total Costs | |
|---|---|---|---|---|---|---|---|---|---|---|---|---|---|---|---|
| Cases | *Peak Warming* | *Peak year* | *Size* | *Duration (year)* | *2015-2100 ($GtCO_2$)* | *2015-2150 ($GtCO_2$)* | *2021-2050* | *2051-2100* | *2101-2150* | *2151-2200* | *$CO_2$* | *$CH_4$* | *$N_2O$* | *Relative* | *Absolute (trillion US$)* |
| i) Reference pathway (without constraints) | 2.30 °C | 2071 | 0.30 °C | 52 | 1,574 | 1,684 | | | | | | | | | 66.1 |
| ii) Illustrative pathway (with constraints) | 2.30 °C | 2070 | 0.30 °C | 53 | 1,553 | 1,670 | -3.5% | 1.8% | 0.4% | -0.2% | 1.9% | -5.7% | -5.4% | 1.1% | 66.8 |
| iii) GWP100 | 2.27 °C | 2072 | 0.27 °C | 50 | 1,497 | 1,548 | 5.1% | 2.5% | -1.5% | -3.1% | 4.7% | -20.9% | -10.6% | 2.1% | 67.5 |
| iv) GWP20 | 2.25 °C | 2072 | 0.25 °C | 50 | 1,687 | 1,803 | 5.2% | 5.2% | 5.2% | 4.7% | -4.9% | 107.6% | -22.3% | 5.2% | 69.5 |
| v) GTP100 | 2.28 °C | 2071 | 0.28 °C | 52 | 1,184 | 1,132 | 31.4% | 12.9% | -0.7% | -4.3% | 23.1% | -86.8% | -4.1% | 12.5% | 74.4 |
| vi) GTP20 | 2.25 °C | 2072 | 0.25 °C | 50 | 1,649 | 1,752 | 4.1% | 3.6% | 3.0% | 2.2% | -3.1% | 71.3% | -15.9% | 3.6% | 68.4 |
| vii) GWP72 (cost-effective time horizon) | 2.26 °C | 2072 | 0.26 °C | 50 | 1,541 | 1,608 | 3.9% | 2.2% | -0.7% | -2.1% | 2.3% | -0.5% | -10.4% | 1.9% | 67.3 |
| viii) GTP33 (cost-effective time horizon) | 2.26 °C | 2072 | 0.26 °C | 50 | 1,542 | 1,610 | 3.9% | 2.2% | -0.7% | -2.1% | 2.3% | -0.6% | -6.3% | 1.9% | 67.3 |
| ix) Best available GWPs (time-dependent) | 2.28 °C | 2072 | 0.28 °C | 51 | 1,590 | 1,667 | 0.9% | 0.7% | 1.0% | -0.6% | -0.4% | 14.0% | -16.2% | 0.7% | 66.5 |
| x) Best available GTPs (time-dependent) | 2.31 °C | 2072 | 0.31 °C | 53 | 1,539 | 1,700 | 1.4% | 0.7% | -1.1% | 1.2% | 1.3% | -4.6% | -12.7% | 0.6% | 66.5 |

**Supplementary Table 2. The characteristics of pathways derived for metric cost calculations.** See the table caption in the last page of this table.

| **c. 2 °C overshoot 2150** | **Peak Profile** | | **Overshoot Profile** | | **Carbon Budget** | | **Cost distribution (period)** | | | | **Cost distribution (gas)** | | | **Total Costs** | |
|---|---|---|---|---|---|---|---|---|---|---|---|---|---|---|---|
| **Cases** | *Peak Warming* | *Peak year* | *Size* | *Duration (year)* | *2015-2100 ($GtCO_2$)* | *2015-2150 ($GtCO_2$)* | *2021-2050* | *2051-2100* | *2101-2150* | *2151-2200* | *$CO_2$* | *$CH_4$* | *$N_2O$* | *Relative* | *Absolute (trillion US$)* |
| i) Reference pathway (without constraints) | 3.05 °C | 2090 | 1.05 °C | 107 | 3,328 | 2,285 | | | | | | | | | 37.4 |
| ii) Illustrative pathway (with constraints) | 2.90 °C | 2086 | 0.90 °C | 107 | 2,901 | 2,028 | 42.9% | 34.0% | -15.0% | -3.5% | 12.9% | -40.4% | 11.5% | 8.0% | 40.4 |
| iii) GWP100 | 2.94 °C | 2091 | 0.94 °C | 105 | 3,340 | 2,203 | 4.3% | 6.6% | 2.2% | -2.5% | 4.2% | 2.3% | -2.4% | 3.9% | 38.9 |
| iv) GWP20 | 2.90 °C | 2090 | 0.90 °C | 105 | 3,465 | 2,438 | -7.2% | 15.4% | 7.7% | 6.0% | -4.5% | 161.5% | -17.9% | 10.5% | 41.3 |
| v) GTP100 | 2.97 °C | 2090 | 0.97 °C | 106 | 3,056 | 1,794 | 28.3% | 19.4% | 6.1% | -4.6% | 21.9% | -82.9% | 5.2% | 12.0% | 41.9 |
| vi) GTP20 | 2.91 °C | 2090 | 0.91 °C | 105 | 3,444 | 2,393 | -3.9% | 11.8% | 5.6% | 3.4% | -3.0% | 117.2% | -10.5% | 7.9% | 40.3 |
| vii) GWP106 (cost-effective time horizon) | 2.94 °C | 2091 | 0.94 °C | 105 | 3,334 | 2,193 | 4.6% | 6.6% | 2.1% | -2.7% | 4.7% | -1.9% | -2.7% | 3.9% | 38.9 |
| viii) GTP39 (cost-effective time horizon) | 2.94 °C | 2091 | 0.94 °C | 105 | 3,332 | 2,190 | 4.8% | 6.7% | 2.1% | -2.8% | 4.7% | -4.4% | 6.3% | 3.9% | 38.9 |
| ix) Best available GWPs (time-dependent) | 2.95 °C | 2091 | 0.95 °C | 105 | 3,385 | 2,317 | -0.3% | 3.3% | 1.4% | 1.1% | -0.7% | 32.3% | -9.6% | 2.2% | 38.2 |
| x) Best available GTPs (time-dependent) | 3.06 °C | 2090 | 1.06 °C | 106 | 3,309 | 2,225 | 1.5% | 1.2% | 0.3% | -1.3% | 2.4% | -14.6% | -7.1% | 0.7% | 37.6 |

| **d. 1.5 °C overshoot 2100** | **Peak Profile** | | **Overshoot Profile** | | **Carbon Budget** | | **Cost distribution (period)** | | | | **Cost distribution (gas)** | | | **Total Costs** | |
|---|---|---|---|---|---|---|---|---|---|---|---|---|---|---|---|
| **Cases** | *Peak Warming* | *Peak year* | *Size* | *Duration (year)* | *2015-2100 ($GtCO_2$)* | *2015-2150 ($GtCO_2$)* | *2021-2050* | *2051-2100* | *2101-2150* | *2151-2200* | *$CO_2$* | *$CH_4$* | *$N_2O$* | *Relative* | *Absolute (trillion US$)* |
| i) Reference pathway (without constraints) | 1.98 °C | 2062 | 0.48 °C | 70 | 616 | 479 | | | | | | | | | 108.3 |
| ii) Illustrative pathway (with constraints) | 1.95 °C | 2058 | 0.45 °C | 72 | 549 | 440 | 12.5% | 3.3% | -0.2% | -0.9% | 6.8% | -17.7% | 11.1% | 4.5% | 113.1 |
| iii) GWP100 | 1.93 °C | 2063 | 0.43 °C | 69 | 544 | 362 | 6.0% | 1.9% | -1.8% | -2.9% | 4.6% | -18.9% | -3.1% | 2.2% | 110.7 |
| iv) GWP20 | 1.90 °C | 2063 | 0.40 °C | 69 | 722 | 605 | 9.4% | 4.7% | 5.1% | 5.1% | -5.2% | 107.7% | -16.8% | 5.6% | 114.3 |
| v) GTP100 | 1.94 °C | 2062 | 0.44 °C | 70 | 237 | -42 | 33.6% | 10.2% | -1.3% | -4.0% | 23.8% | -85.6% | 7.1% | 13.0% | 122.3 |
| vi) GTP20 | 1.91 °C | 2063 | 0.41 °C | 69 | 687 | 557 | 6.9% | 3.3% | 2.8% | 2.6% | -3.4% | 72.0% | -9.9% | 3.8% | 112.4 |
| vii) GWP74 (cost-effective time horizon) | 1.92 °C | 2063 | 0.42 °C | 69 | 583 | 415 | 5.1% | 1.7% | -1.1% | -1.9% | 2.4% | -0.4% | -3.2% | 2.0% | 110.4 |
| viii) GTP34 (cost-effective time horizon) | 1.92 °C | 2063 | 0.42 °C | 69 | 587 | 421 | 5.0% | 1.7% | -1.0% | -1.8% | 2.1% | 1.2% | 1.1% | 2.0% | 110.4 |
| ix) Best available GWPs (time-dependent) | 1.94 °C | 2063 | 0.44 °C | 70 | 634 | 473 | 1.4% | 0.6% | 0.9% | -0.3% | -0.6% | 14.5% | -10.7% | 0.7% | 109.1 |
| x) Best available GTPs (time-dependent) | 1.98 °C | 2063 | 0.48 °C | 70 | 585 | 495 | 1.4% | 0.5% | -0.5% | 1.2% | 1.4% | -5.7% | -7.0% | 0.6% | 108.9 |

**Supplementary Table 2 (Cont.) | The characteristics of pathways derived for metric cost calculations.** See the table caption in the last page of this table.

| e. 1.5 °C overshoot 2150 | Peak Profile | | Overshoot Profile | | Carbon Budget | | Cost distribution (period) | | | | Cost distribution (gas) | | | Total Costs | |
|---|---|---|---|---|---|---|---|---|---|---|---|---|---|---|---|
| Cases | *Peak Warming* | *Peak year* | *Size* | *Duration (year)* | *2015-2100 ($GtCO_2$)* | *2015-2150 ($GtCO_2$)* | *2021-2050* | *2051-2100* | *2101-2150* | *2151-2200* | *$CO_2$* | *$CH_4$* | *$N_2O$* | *Relative* | *Absolute (trillion US$)* |
| i) Reference pathway (without constraints) | 2.76 °C | 2083 | 1.26 °C | 122 | 2,533 | 1,242 | | | | | | | | | 55.3 |
| ii) Illustrative pathway (with constraints) | 2.48 °C | 2075 | 0.98 °C | 122 | 1,759 | 836 | 104.1% | 55.9% | -29.5% | -6.8% | 24.5% | -46.2% | 20.1% | 17.9% | 65.2 |
| iii) GWP100 | 2.64 °C | 2083 | 1.14 °C | 121 | 2,545 | 1,169 | 7.1% | 5.8% | 2.2% | -2.7% | 3.9% | 4.6% | 3.3% | 4.0% | 57.6 |
| iv) GWP20 | 2.60 °C | 2082 | 1.10 °C | 121 | 2,665 | 1,398 | 6.6% | 13.9% | 8.4% | 6.5% | -4.9% | 167.0% | -12.4% | 10.9% | 61.4 |
| v) GTP100 | 2.67 °C | 2082 | 1.17 °C | 122 | 2,260 | 764 | 29.7% | 16.4% | 6.3% | -5.2% | 21.6% | -82.0% | 11.5% | 11.8% | 61.9 |
| vi) GTP20 | 2.61 °C | 2082 | 1.11 °C | 121 | 2,646 | 1,354 | 6.3% | 10.6% | 6.2% | 3.7% | -3.3% | 121.7% | -4.7% | 8.2% | 59.9 |
| vii) GWP110 (cost-effective time horizon) | 2.64 °C | 2083 | 1.14 °C | 121 | 2,535 | 1,152 | 7.5% | 5.8% | 2.1% | -3.0% | 4.6% | -2.3% | 2.8% | 4.0% | 57.5 |
| viii) GTP40(cost-effective time horizon) | 2.64 °C | 2083 | 1.14 °C | 121 | 2,538 | 1,156 | 7.4% | 5.8% | 2.1% | -3.0% | 4.4% | -2.2% | 12.5% | 4.0% | 57.5 |
| ix) Best available GWPs (time-dependent) | 2.65 °C | 2083 | 1.15 °C | 121 | 2,590 | 1,277 | 2.4% | 3.0% | 1.6% | 1.1% | -0.8% | 33.6% | -4.2% | 2.3% | 56.6 |
| x) Best available GTPs (time-dependent) | 2.76 °C | 2084 | 1.26 °C | 122 | 2,514 | 1,185 | 1.7% | 1.1% | 0.3% | -1.2% | 2.3% | -14.8% | -2.3% | 0.7% | 55.7 |

**Supplementary Table 2 (Cont.) | The characteristics of pathways derived for metric cost calculations.** The table shows the temperature peak and overshoot profiles, carbon budget, and total mitigation costs and their distributions across four periods and across gases. Panels **a** to **e** correspond to the five main pathways considered in this study. Case i shows the reference pathway, against which the additional costs of using metrics in relative terms in percent (i.e. Cases iii to x) are calculated. In comparison, Case ii is the illustrative pathway in Fig. 2 of the main paper. The illustrative pathway, which is obtained under the abatement constraints, is not identical to the reference pathway, which does not use such constraints (Methods). Case ii is shown only for comparison with Case i and not used for metric cost calculations. Cases iii to vi show the pathways using GWP100, GWP20, GTP100, and GTP20, respectively. Cases vii and viii are those applying the optimal cost-effective time horizon for GWP and GTP, respectively. In Cases ix and x, pathways using best available GWPs and GTPs (selected from the default set of three metrics), respectively, are presented. The overshoot size refers to the maximum warming relative to the respective stabilization target. The carbon budgets presented above account for only $CO_2$ (i.e. non-$CO_2$ components not included) and indicate net emissions. Absolute costs are in net present values in trillion US$2010. See also Supplementary Figs. 12 to 16 for further details in the results presented here.

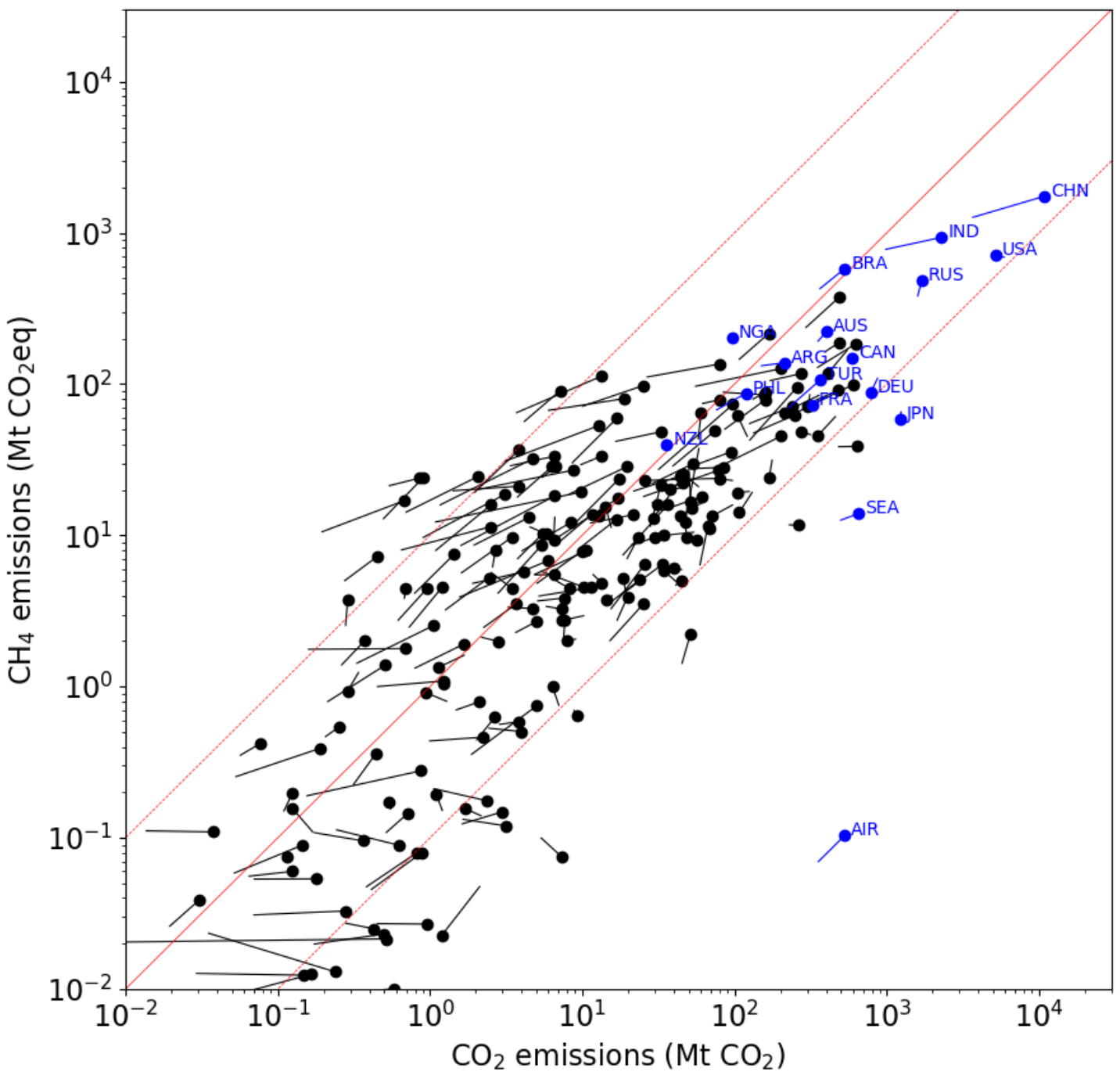

**Supplementary Fig. 1. Scatter plot of $CH_4$ emissions (as $MtCO_2eq$/year using GWP100) versus $CO_2$ emissions (as $MtCO_2$/year) on a country-by-country basis.** The segments indicate the time evolution from year 2000 to 2015 (with the filled circles indicating the latest date) using version 5.0 of the Emission Database for Global Atmospheric Research (EDGAR) (Crippa et al., 2020, *Scientific Data*). Emissions from international maritime and aircraft transport are also shown. For largest emitters like China, India, and the US, $CO_2$ emissions are by far larger than $CH_4$ emissions (via GWP100). $CH_4$ emissions have a higher significance in countries including but not limited to Argentina, Australia, Brazil, Nigeria, New Zealand, and the Philippines. Countries listed here are non-exhaustive. Note the log-log scale. Acronyms are those used in the EDGAR database: AIR: international aviation, ARG: Argentina, AUS: Australia, BRA: Brazil, CAN: Canada, CHN: China, DEU: Germany, FRA: France, IND: India, JPN: Japan, NGA: Nigeria, NZL: New-Zealand, PHL: Philippines, RUS: Russia, TUR: Turkey, USA: United States, SEA: international shipping.

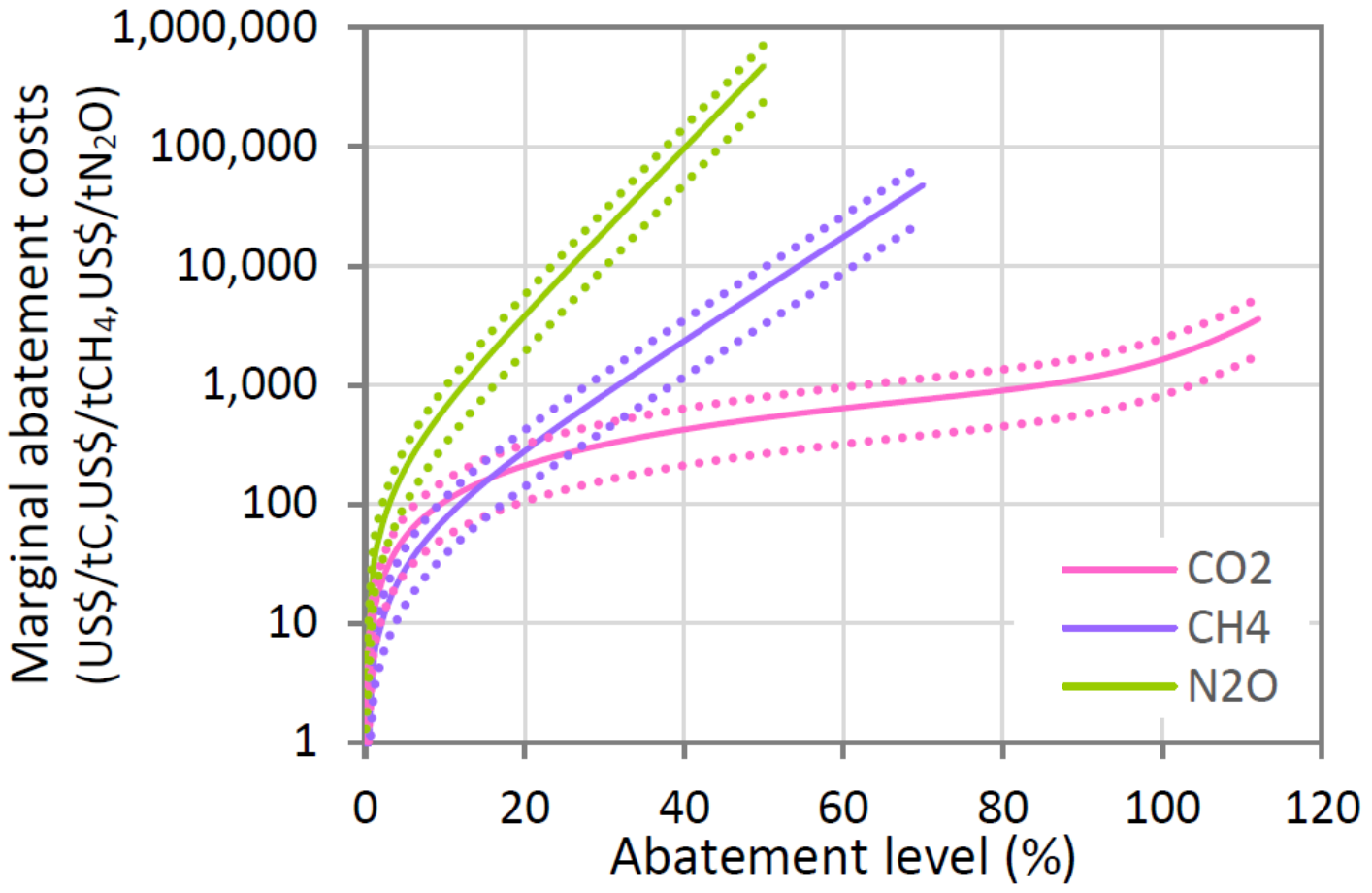

**Supplementary Fig. 2. The marginal abatement cost (MAC) curves for $CO_2$ (fossil fuel origin), $CH_4$, and $N_2O$.** Solid lines indicate the MAC curves used in our analysis under default assumptions. Different colors are assigned for different gases as indicated in the legend. The MAC curves are given as a function of the abatement level relative to the baseline level (Supplementary Fig. 3a,b). The maximum abatement level for $CO_2$, $CH_4$, and $N_2O$ is assumed at 112%, 70%, and 50%, respectively. Uncertainty ranges of ±50% are assumed for the three MAC curves as shown by dotted lines. The logarithmic scale is used on the vertical axis.

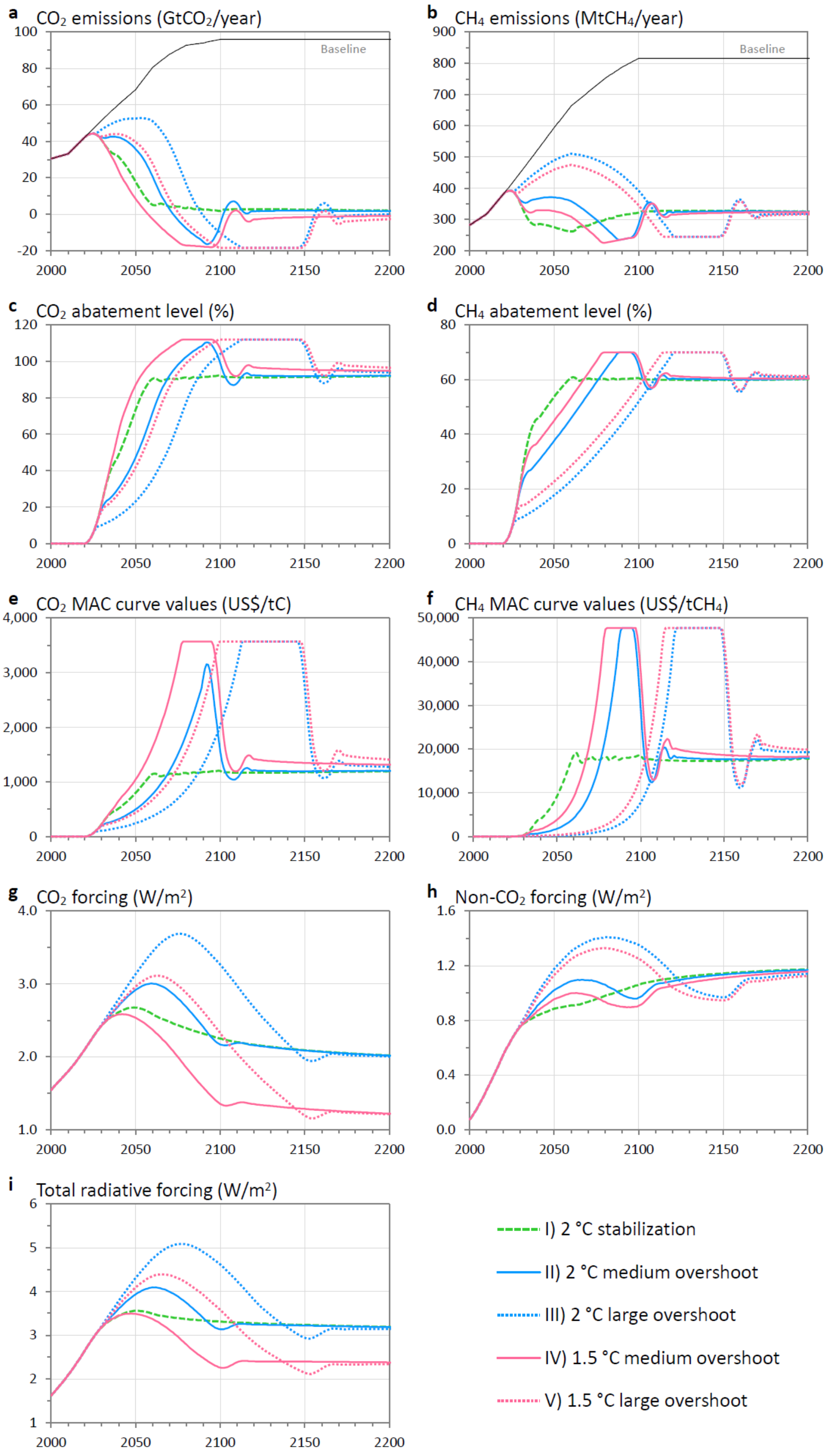

**Supplementary Fig. 3. Details of the five illustrative pathways presented in Fig. 2 of the main paper.** Panels **a** and **b** show the global total anthropogenic emissions of $CO_2$ and $CH_4$, respectively. Panels **c** and **d** indicate the abatement levels of $CO_2$ and $CH_4$, respectively, relative to assumed baseline levels (black lines in panels **a** and **b**). Panels **e** and **f** present the values of the $CO_2$ and $CH_4$ MAC curves, respectively. Note that these MAC curve values are not directly used to calculate GCPs (Methods). Panels **g** and **h** show the radiative forcing of $CO_2$ and non-$CO_2$ climate forcers, respectively. The sum of these two forcing terms gives the total forcing in panel **i**. All panels follow the legend placed at the bottom right of the figure.

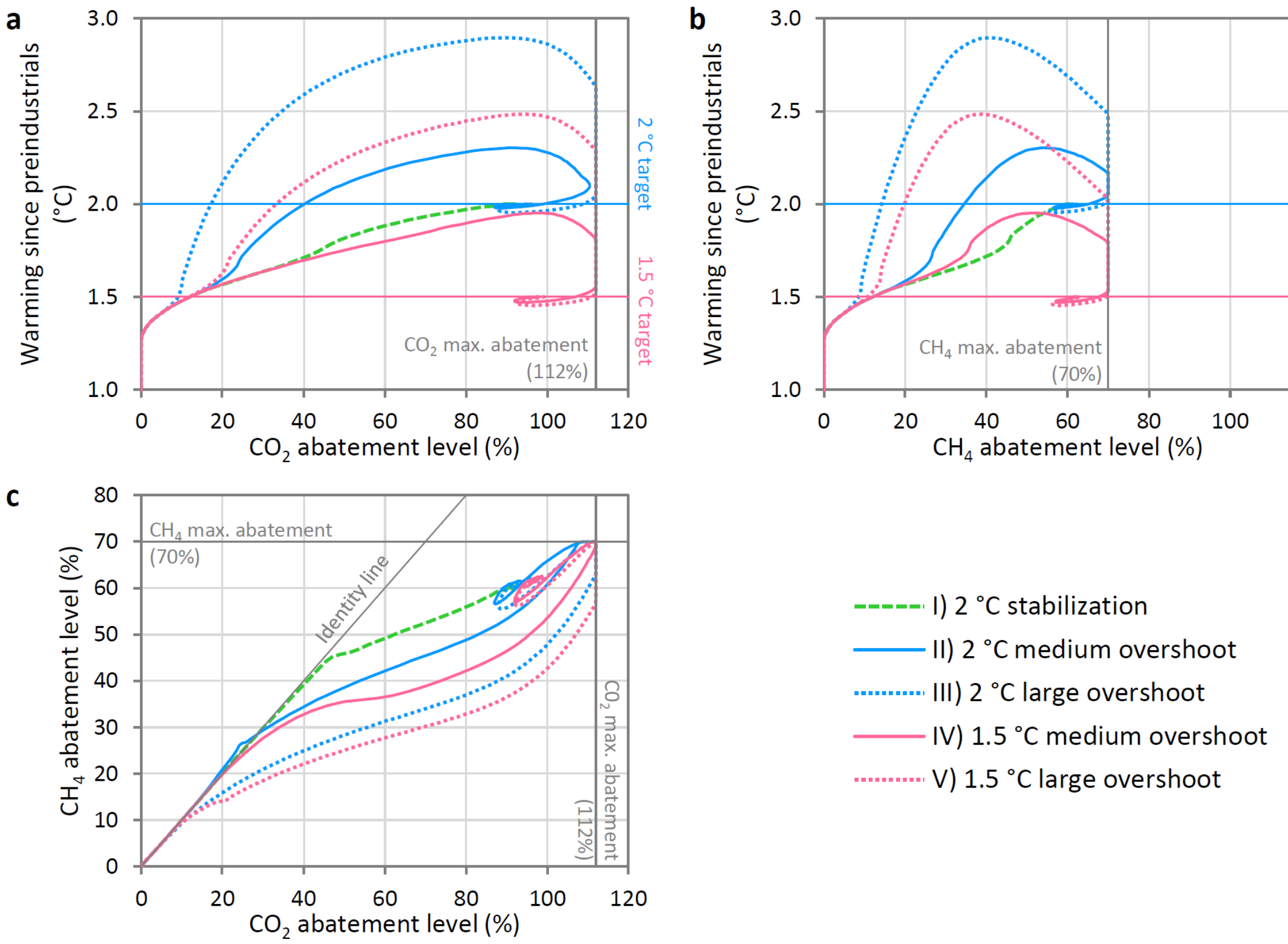

**Supplementary Fig. 4. $CO_2$ and $CH_4$ abatement and warming levels.** Panels **a** and **b** show the relationships between abatement levels of $CO_2$ and $CH_4$, respectively, relative to assumed baseline levels and warming levels along the five illustrative pathways presented in Fig. 2 of the main paper. The 2 and 1.5 °C targets are indicated by thin blue and pink lines, and the $CO_2$ and $CH_4$ maximum abatement levels assumed in our model are in thin grey lines. Panel **c** compares the abatement levels of $CO_2$ and $CH_4$ along the five pathways. The $CO_2$ and $CH_4$ maximum abatement levels as well as the identity line are shown by thin grey lines. All panels follow the legend placed at the bottom left of the figure.

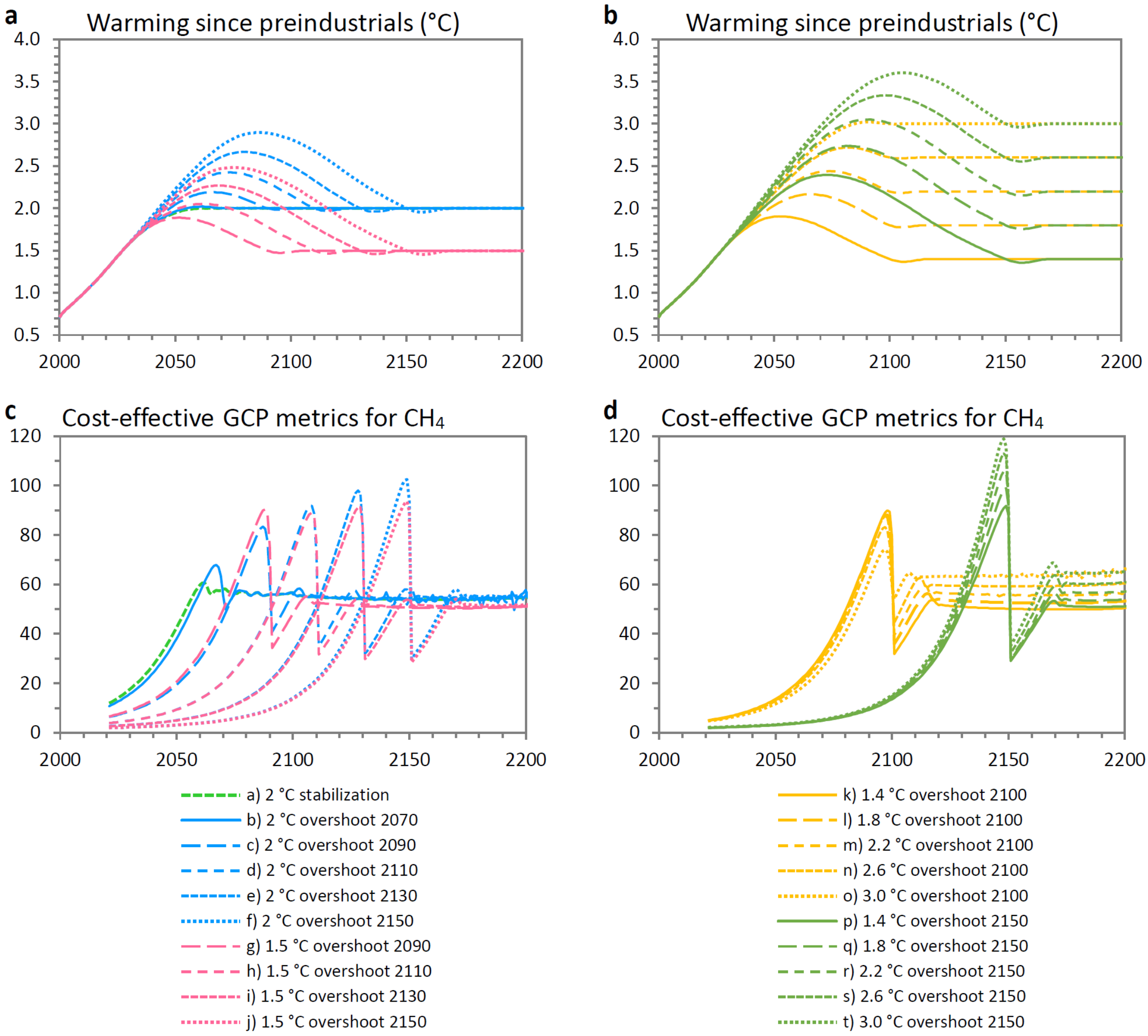

**Supplementary Fig. 5. Sensitivity analysis of the temperature and GCP profiles to the overshoot length and magnitude.** This figure presents the results of a sensitivity analysis of the outcome shown in Fig. 2 of the main paper. Panel **a** shows the sensitivity of the temperature profile to the changes in the overshoot termination period (or the stabilization period) from 2070 to 2150 with a 20-year interval for the 2 and 1.5 °C targets (Cases a to j). There is no feasible pathway for the 1.5 °C target if overshoot is allowed only till 2070. In this panel, the 2 °C non-overshoot case is also shown for comparison. Panel **b** presents the sensitivity of the temperature profile to the changes in the stabilization target level from 1.4 to 3.0 °C with an interval of 0.4 °C when the overshoot is allowed till 2100 or 2150 (Cases k to t). Panels **c** and **d** show the GCP profiles for the cases in panels **a** and **b**, respectively. Panels **a** and **c** supports the finding of the main paper that the temporal profile of GCP is largely determined by the period remaining before the temperature target is met. This finding is further confirmed by panels **b** and **d**, although some changes in the peak value of GCP to the changes in the target level are found.

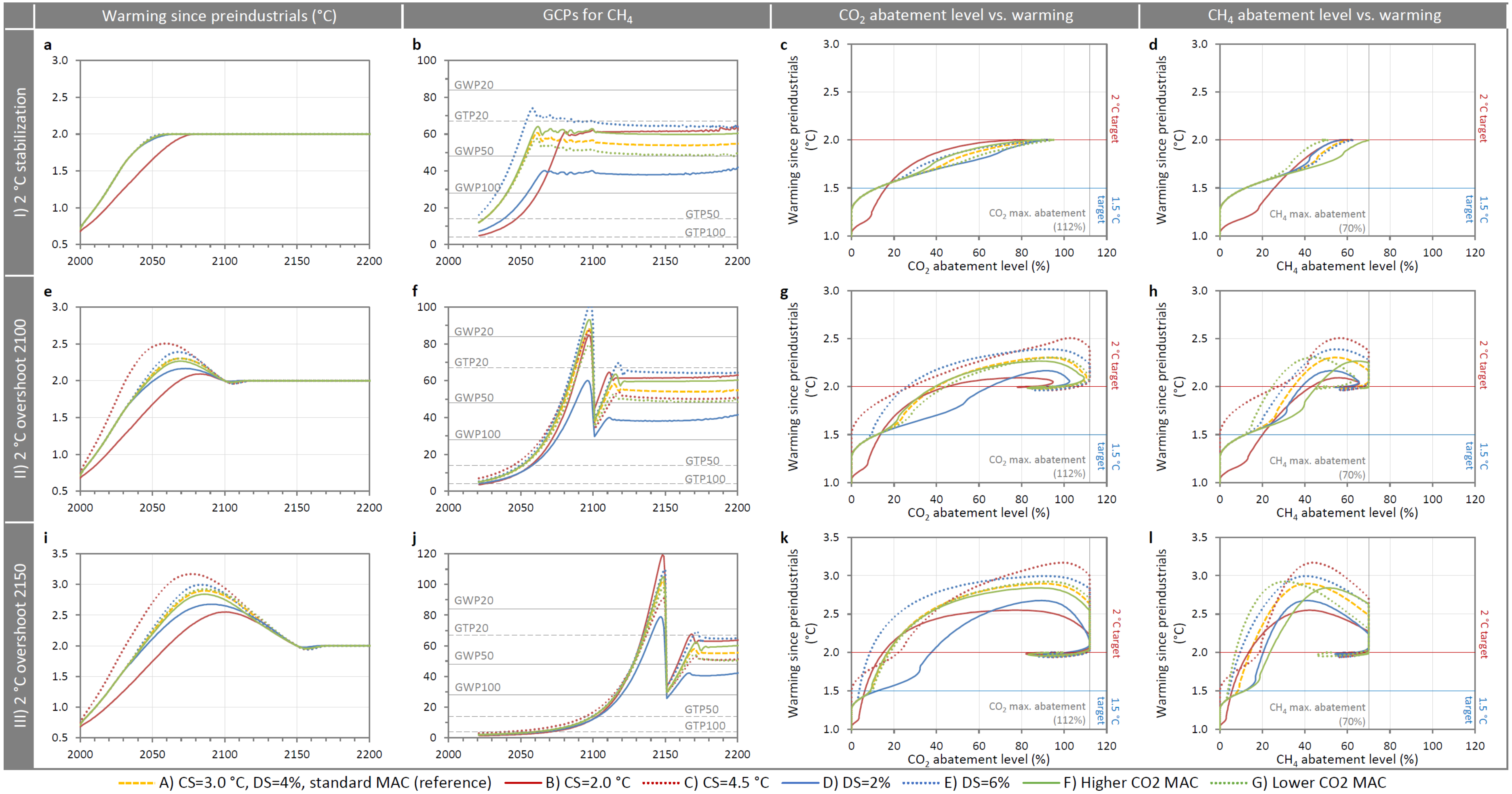

**Supplementary Fig. 6. Temperature pathways, $CH_4$ GCPs, and $CO_2$ and $CH_4$ abatement levels and their sensitivities to the assumptions on the equilibrium climate sensitivity, the discount rate, and the MAC curves.** See the figure caption in the next page.

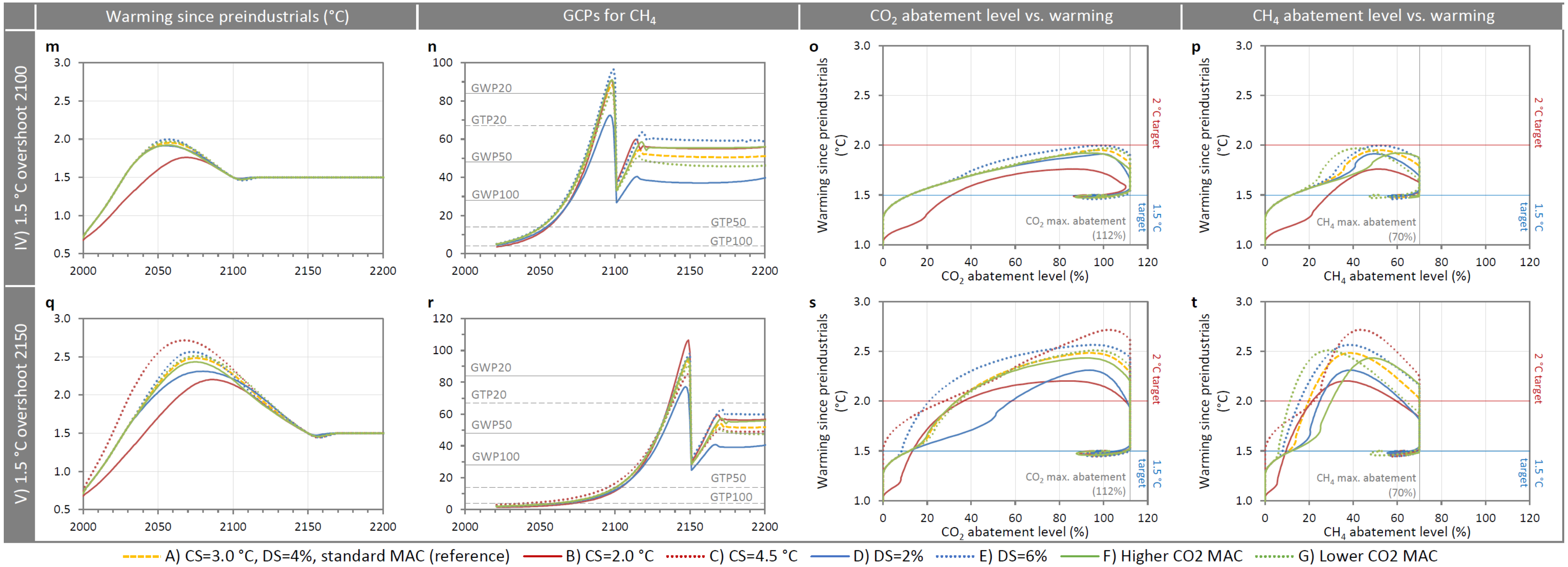

**Supplementary Fig. 6 (Cont.). | Temperature pathways, $CH_4$ GCPs, and $CO_2$ and $CH_4$ abatement levels and their sensitivities to the assumptions on the equilibrium climate sensitivity, the discount rate, and the MAC curves.** The panels on the two columns from left present the results of a sensitivity analysis for the outcome shown in Fig. 2 of the main paper. Those on the two columns from right show the sensitivity results for Supplementary Fig. 4a,b. All panels follow the legend placed at the bottom of the figure. Case A is the reference case using the set of default assumptions. The equilibrium climate sensitivity is set at 3.0 °C in default. In Cases B and C, the climate sensitivity is changed to 2.0 °C and 4.5 °C, respectively. Note that Case C does not exist for the 2 °C stabilization pathway and the 1.5 °C overshoot 2100 pathway because no feasible pathway was found for these targets with the climate sensitivity of 4.5 °C under our model assumptions. In Case D, the discount rate was changed from 4% to 2%, with all other assumptions unchanged. Likewise, in Case E, the discount rate was changed to 6%, with everything else intact. In Case F, while keeping the discount rate at 4%, the $CO_2$ MAC curve was assumed higher by 50% than that in the reference case and the $CH_4$ and $N_2O$ MAC curves were assumed lower than those in the reference case (Supplementary Fig. 2). Case G is an opposite case of Case F, with changes in the assumptions on the MAC curves to opposite directions. For further details of the sensitivity analysis, see Methods of the main paper.

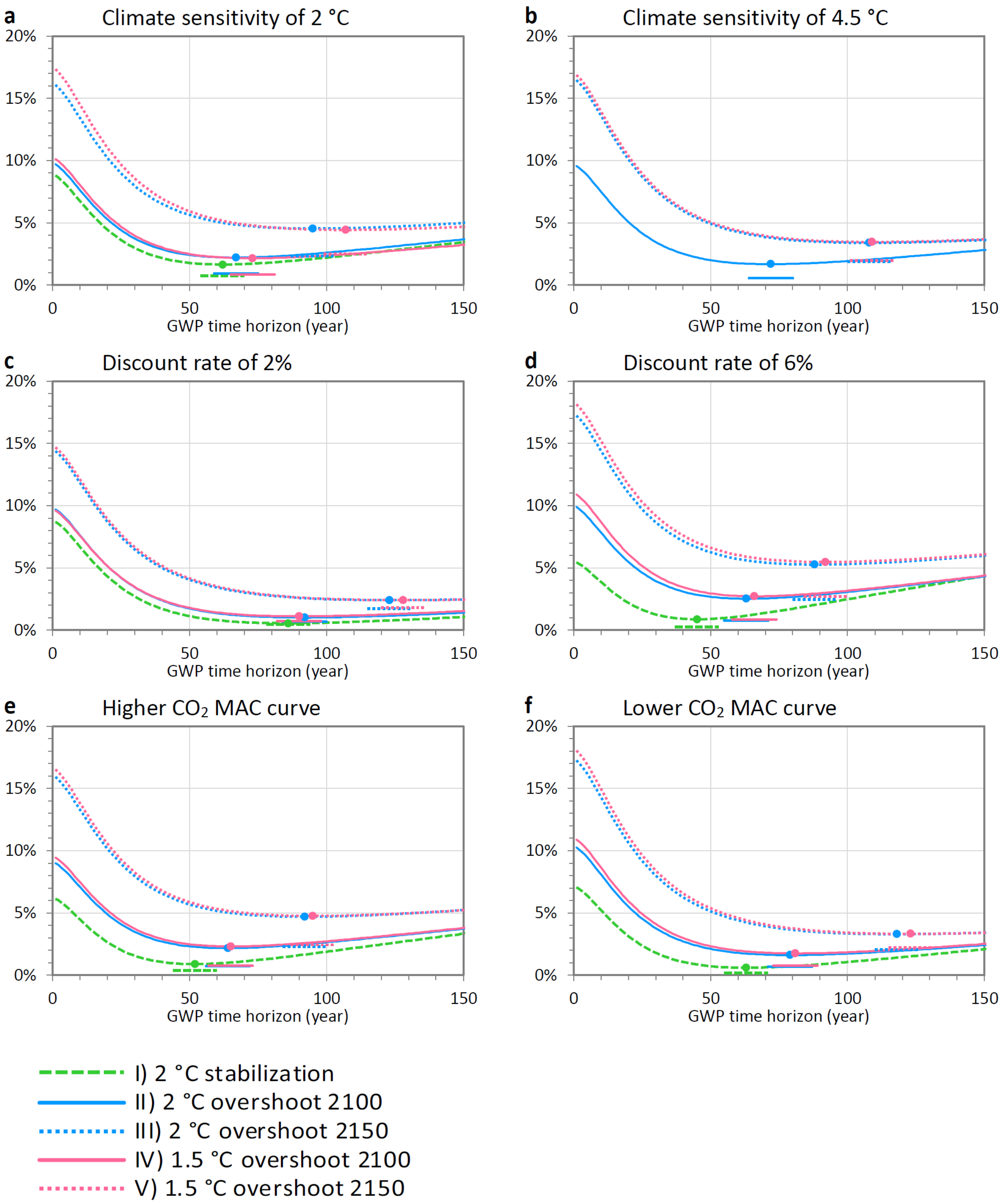

**Supplementary Fig. 7. Sensitivity analysis of the additional mitigation costs of using GWP with a range of time horizons.** The figure indicates the additional total costs with the use of GWP with a time horizon ranging from one to 150 year(s) (relative to the lowest costs without the use of metrics). The figure shows the sensitivity results of Fig. 3a of the main paper. For details of the sensitivity analysis, see Methods of the main paper. All panels follow the legend placed at the bottom left of the figure. The minimum under each pathway, which is marked by a filled circle, indicates the cost-effective time horizon and the associated additional mitigation costs. The horizontal bars vertically aligned with each minimum point indicates the additional mitigation costs of using best available GWPs (from the default set of three metrics GWP100, GWP50, and GWP20) under each pathway. The horizontal bars also follow the legend at the bottom left of the figure. In panel **b**, the 2 °C stabilization pathway and the 1.5 °C medium overshoot pathway are not included in the analysis (Methods of the main paper).

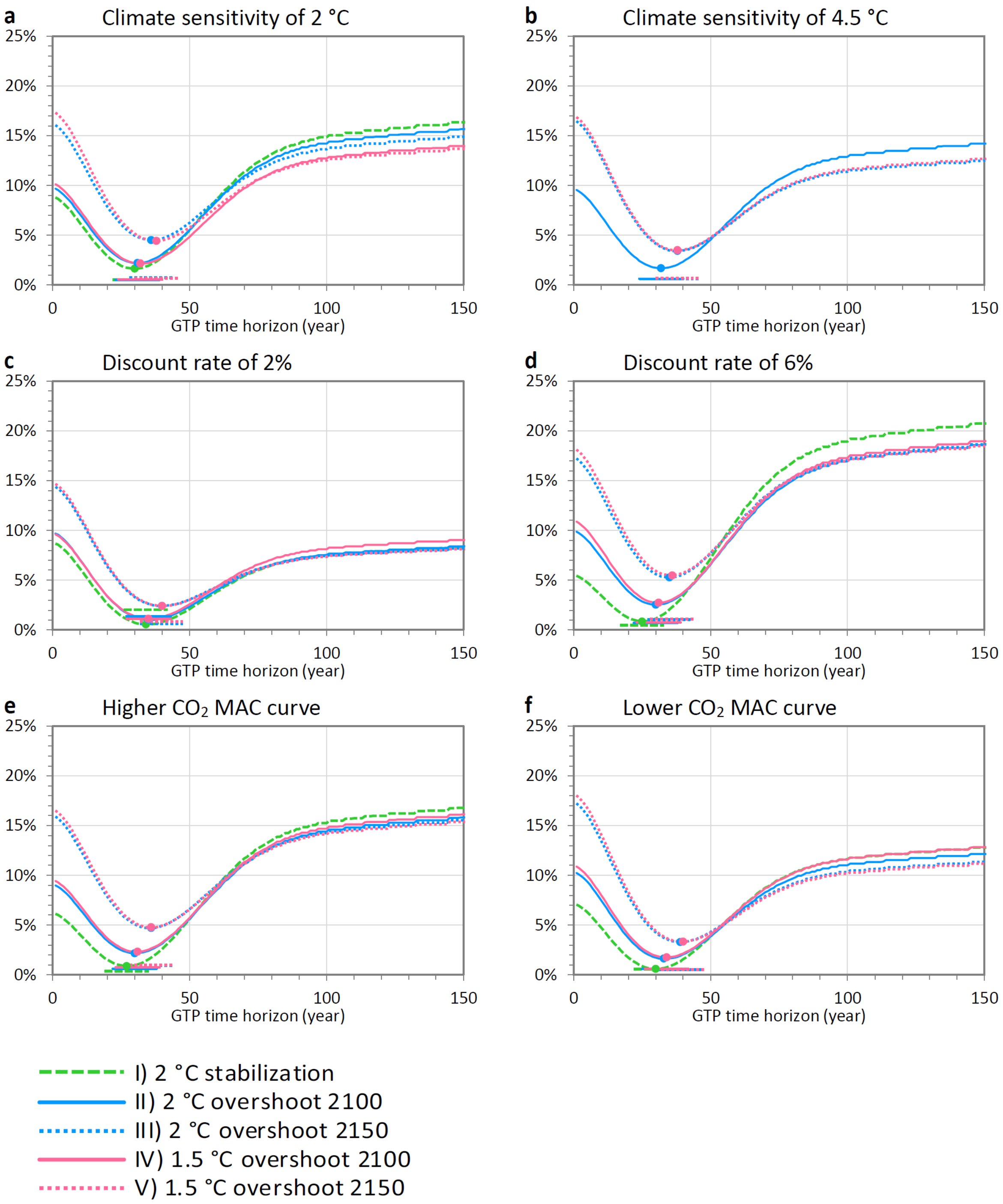

**Supplementary Fig. 8. Sensitivity analysis of the additional mitigation costs of using GTP with a range of time horizons.** The figure presents the sensitivity results of Fig. 3b of the main paper. All panels are presented in the same way with those in Supplementary Fig. 7. See the figure caption for Supplementary Fig. 7.

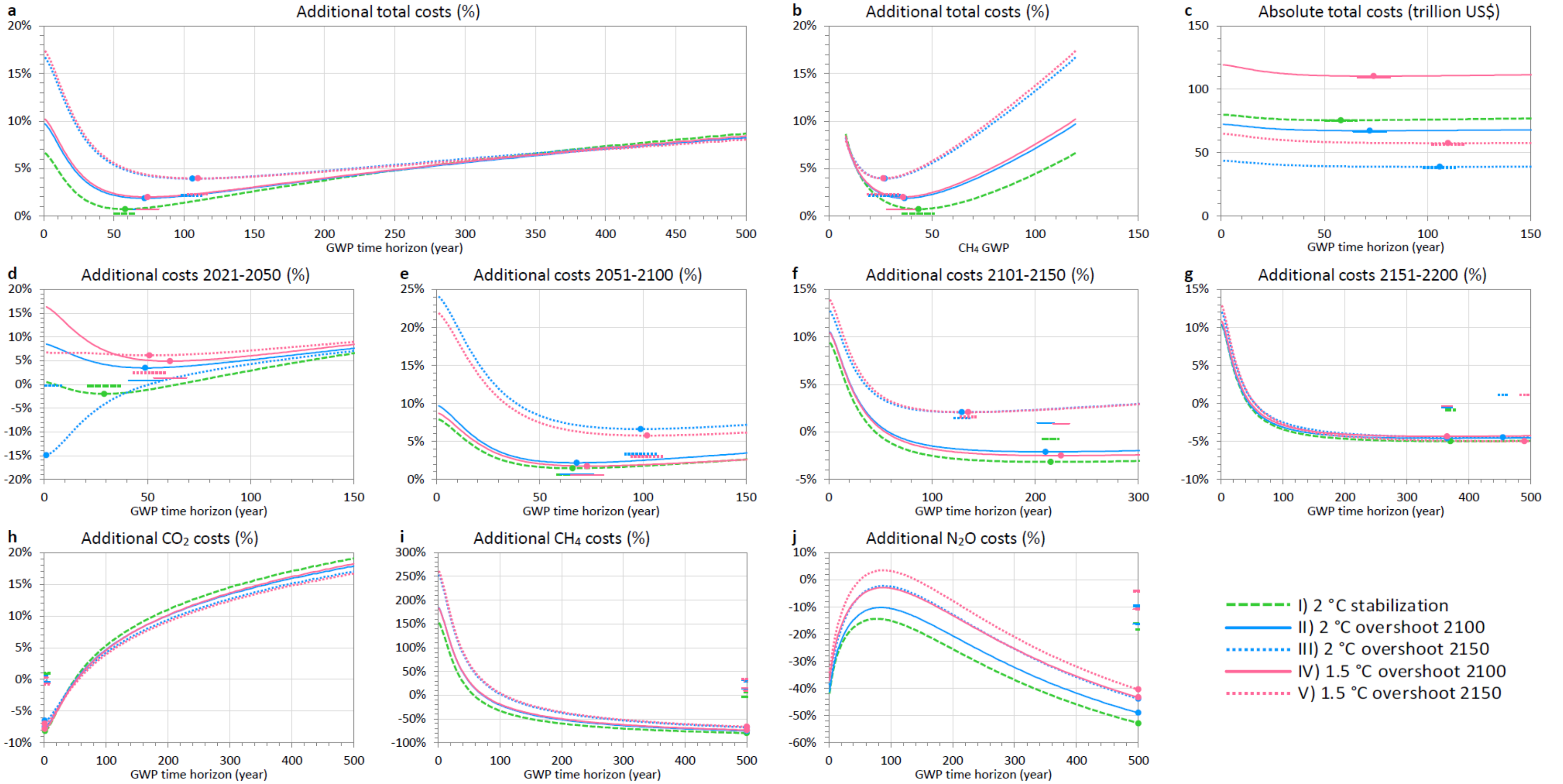

**Supplementary Fig. 9**. **The costs of using GWP and their distributions over time and across gases.** This figure presents further details of the results in Fig. 3a of the main paper. All panels follow the legend placed at the bottom right corner of the figure. In each panel, the minimum under each pathway is marked by a filled circle. The horizontal bars vertically aligned with each minimum point indicate the corresponding costs of using best available GWPs under each pathway. Panel **a** shows the additional mitigation costs of using GWP with a time horizon of up to 500 years. In panel **b**, the same results are presented with metric values on x-axis. Panel **c** shows the absolute mitigation costs (net present value in US$2010) with the use of different GWPs. Panels **d** to **g** show the cost distributions over four periods. The cost distributions over different gases are in panels **h** to **j**.

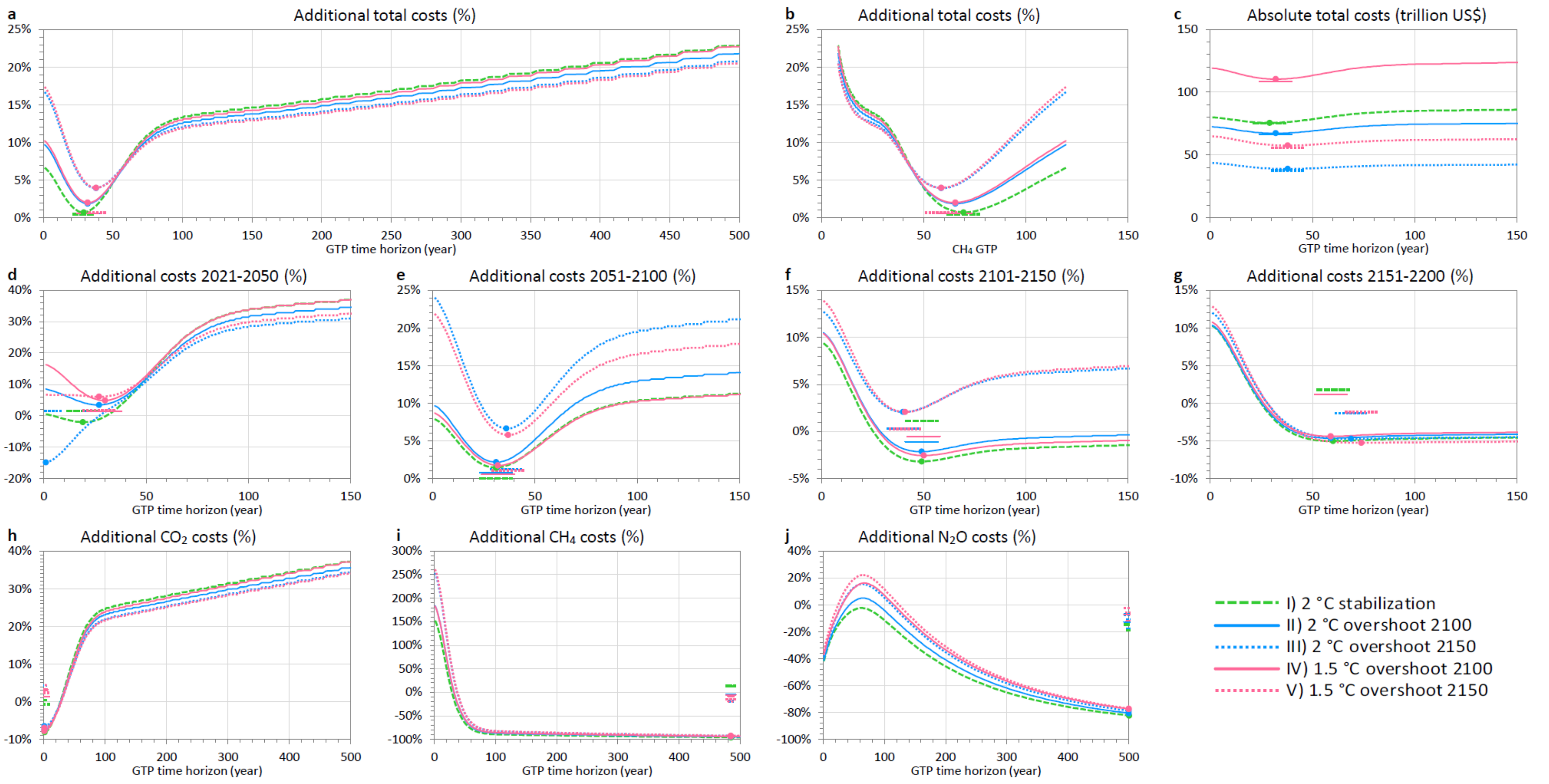

**Supplementary Fig. 10**. **The costs of using GTP and their distributions over time and across gases.** This figure presents further details of the results in Fig. 3b of the main paper. All panels are presented in the same way with those in Supplementary Fig. 9. See the figure caption for Supplementary Fig. 9.

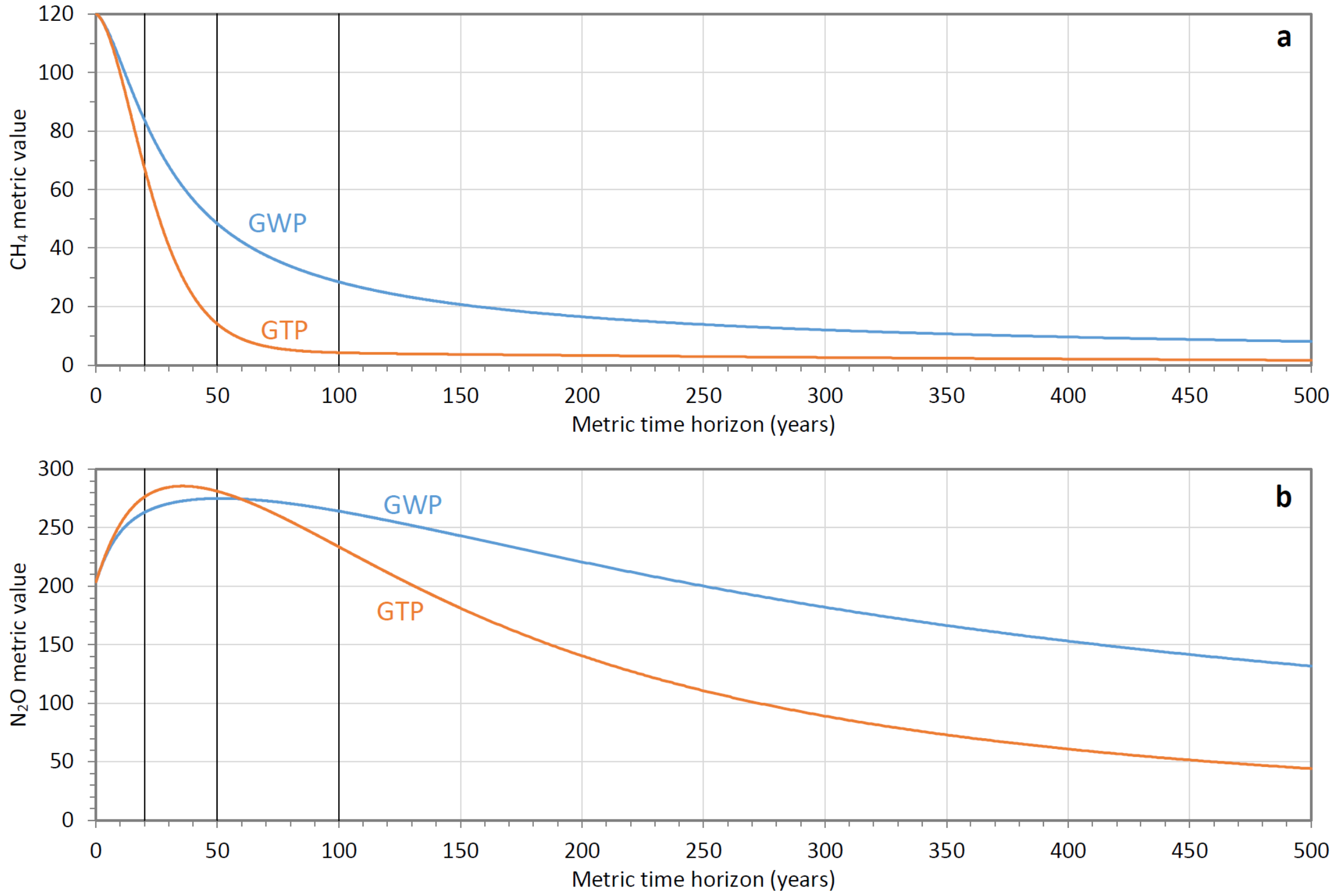

**Supplementary Fig. 11. GWP and GTP values for $CH_4$ and $N_2O$ as a function of time horizon**. Panel **a** shows the GWP and GTP values for $CH_4$ with a time horizon from one to 500 year(s). Those for $N_2O$ are in panel **b**. Representative time horizons (20, 50, and 100 years) are indicated in vertical black lines. The calculation follows Section 8.SM.11 of IPCC AR5 (for metrics *without* inclusion of climate-carbon feedbacks for non-$CO_2$). This figure can be used to clarify the results shown in Fig. 3 of the main paper. The range of time horizons between one and 150 years shown in Fig. 3 corresponds to the $CH_4$ GWP range of 120 to 21 and the $CH_4$ GTP range of 120 to 3.7. The spread of the optimal time horizons for GWP and GTP (Fig. 3 of the main paper) falls on the nearly equivalent metric range between 26 and 43.

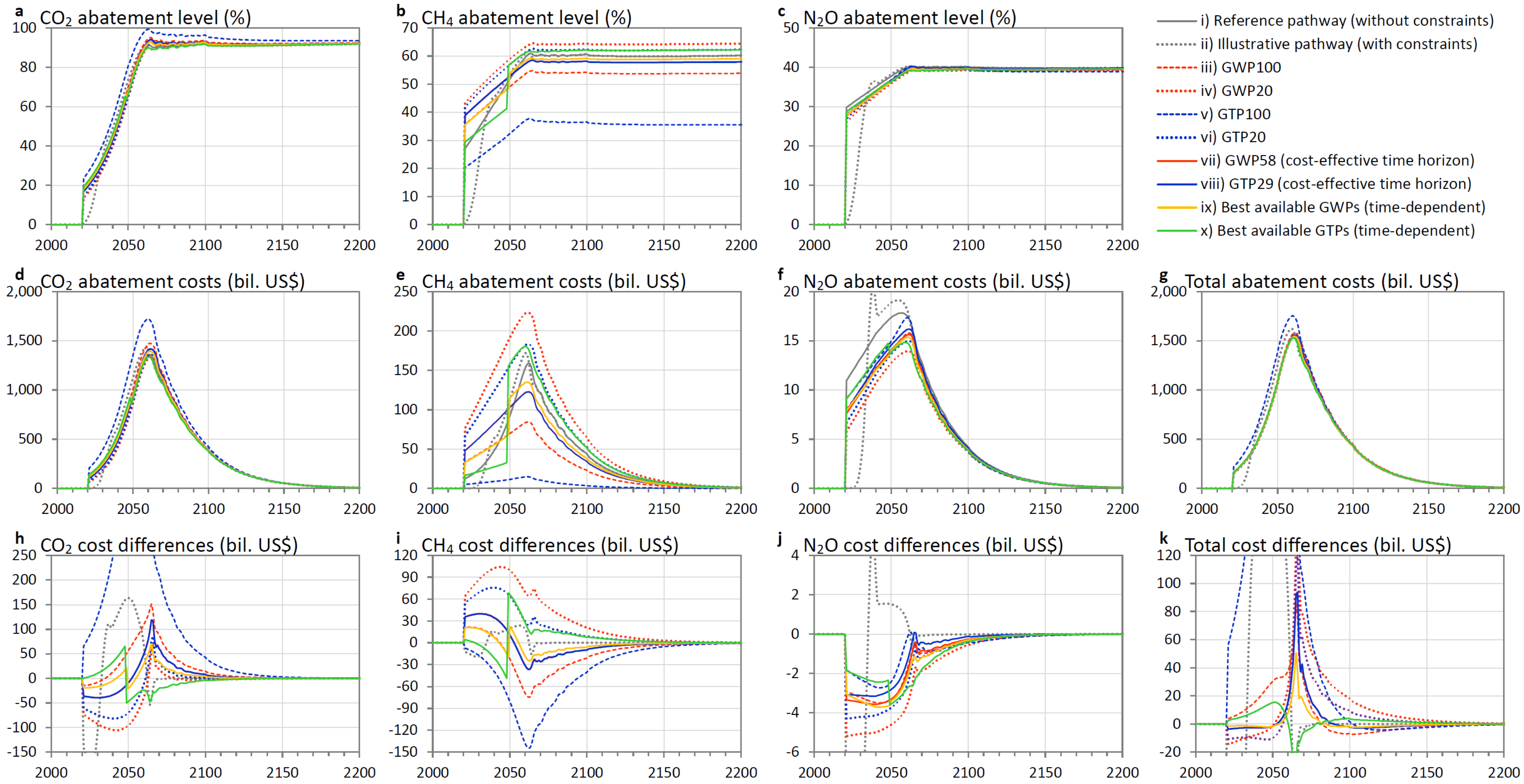

**Supplementary Fig. 12. Complete reporting of the outcomes for metric cost calculations under the 2 °C stabilization pathway.** See the figure caption in the next page.

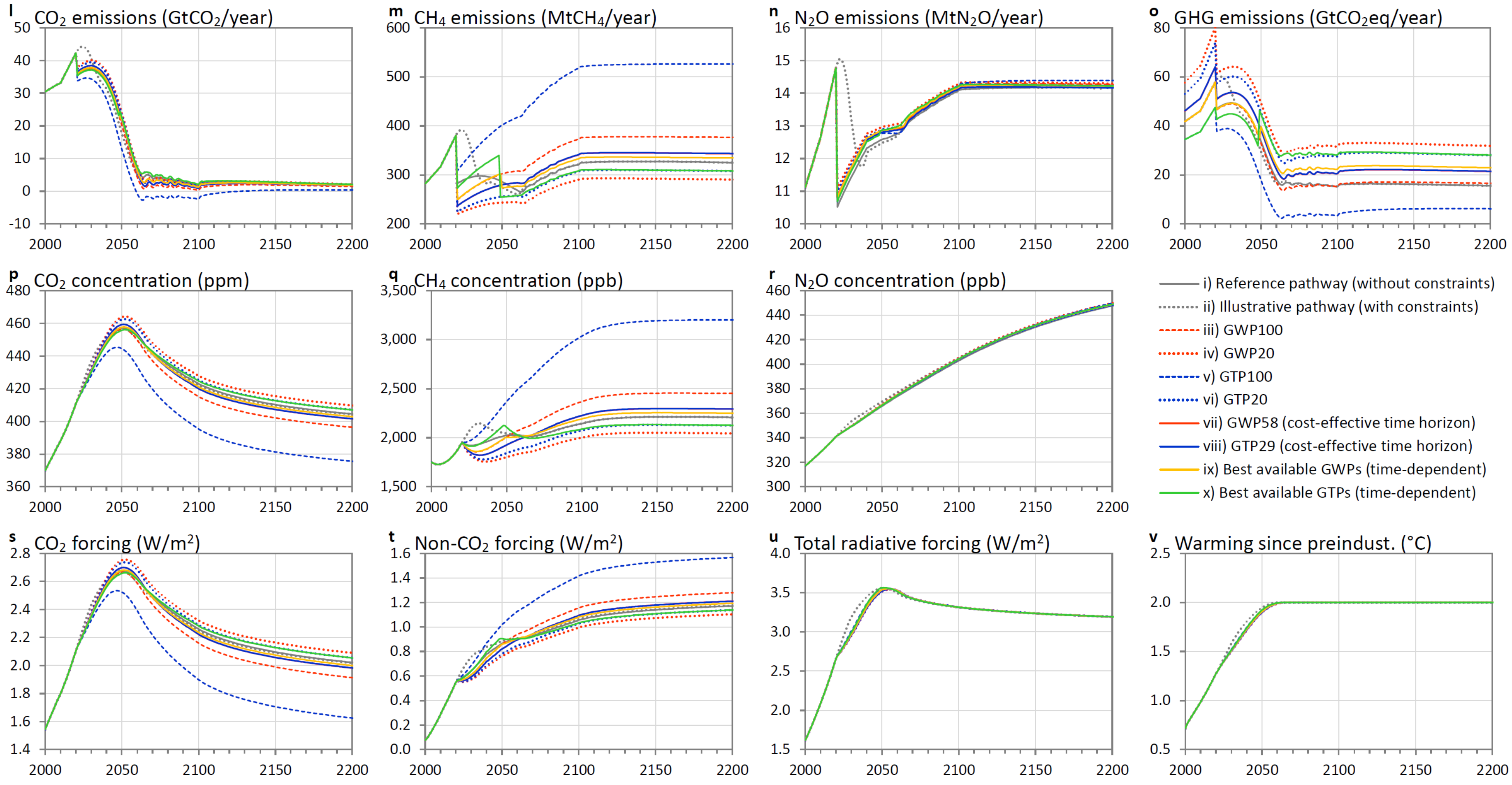

**Supplementary Fig. 12 (Cont.) | Complete reporting of the outcomes for metric cost calculations under the 2 °C stabilization pathway.** All panels follow the legend in each page. Case i shows the reference pathway, against which the additional costs of using metrics (i.e. Cases iii to x) are calculated. In comparison, Case ii presents the illustrative pathway in Fig. 2 of the main paper. The illustrative pathway, which is obtained under the abatement constraints, is not identical to the reference pathway, which does not use such constraints (Methods). Cases iii to vi show the pathways using GWP100, GWP20, GTP100, and GTP20, respectively. Cases vii and viii are those applying the optimal cost-effective time horizon for GWP and GTP, respectively. In Cases ix and x, pathways using best available GWPs and GTPs, respectively, are presented. Costs are in net present values in billion US$2010. In panel **k**, respective metrics are used to aggregate Kyoto gas emissions.

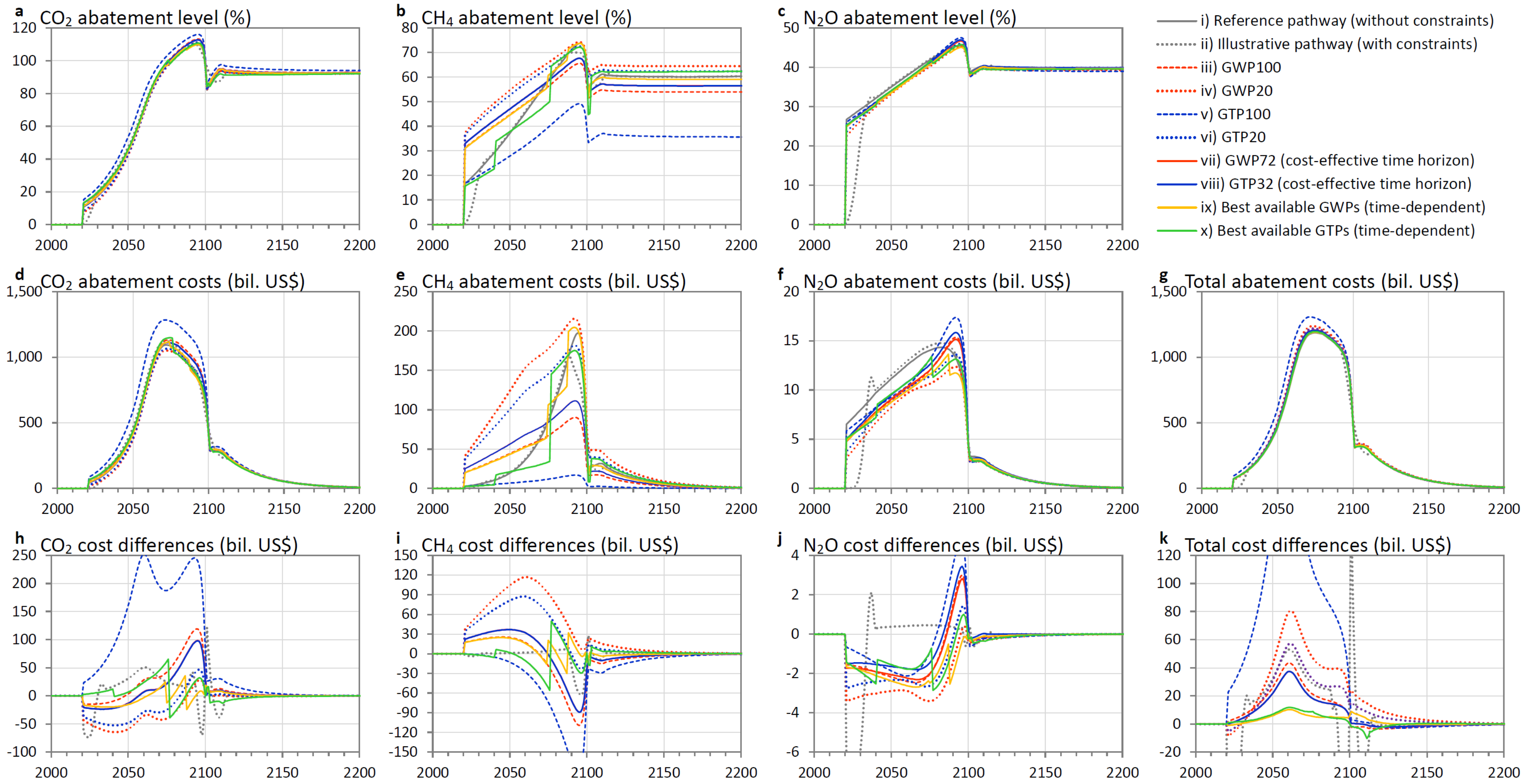

**Supplementary Fig. 13. Complete reporting of the outcomes for metric cost calculations under the 2 °C medium overshoot pathway.** All panels are presented in the same way with those in Supplementary Fig. 12.

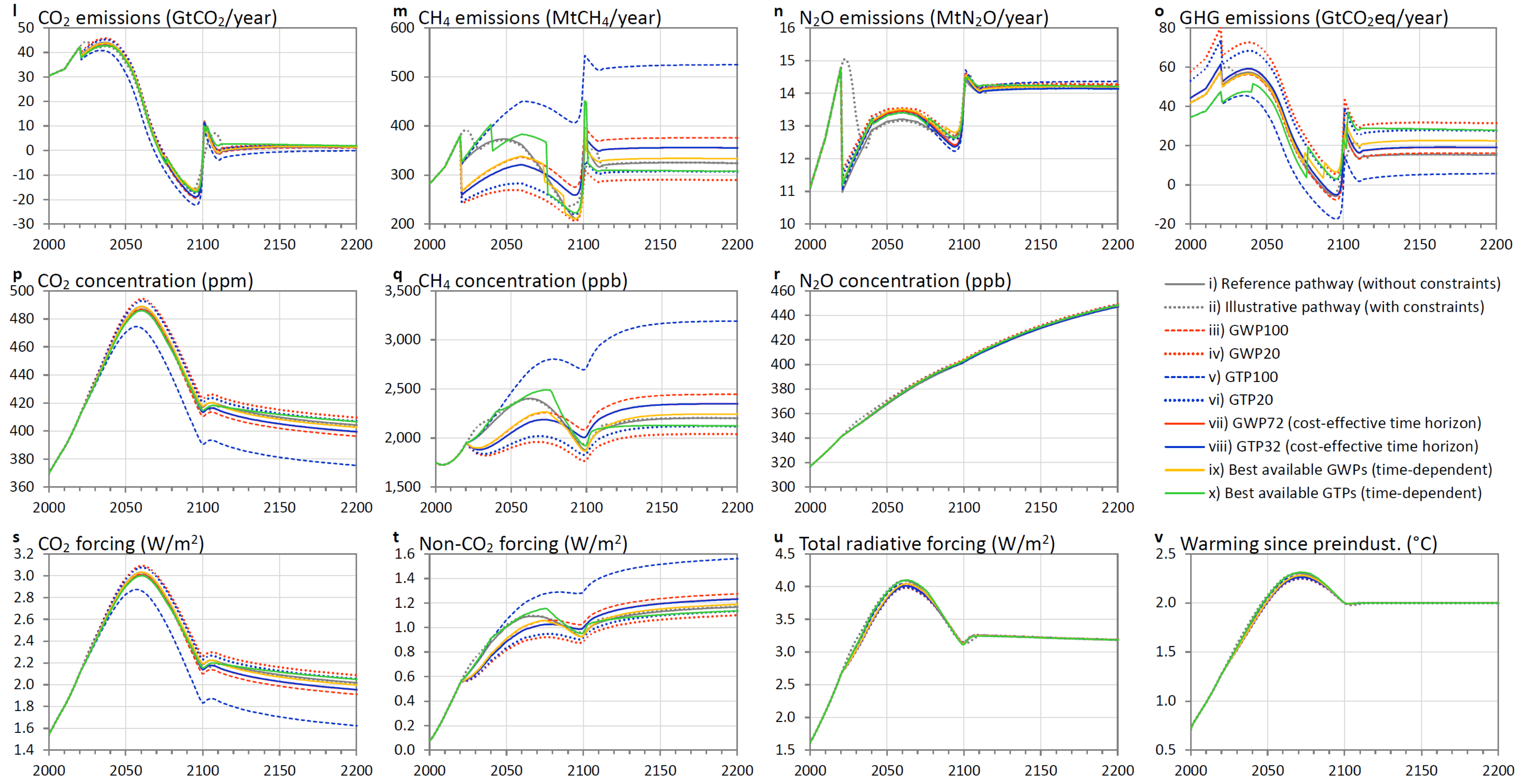

**Supplementary Fig. 13 (Cont.) | Complete reporting of the outcomes for metric cost calculations under the 2 °C medium overshoot pathway.** All panels are presented in the same way with those in Supplementary Fig. 12.

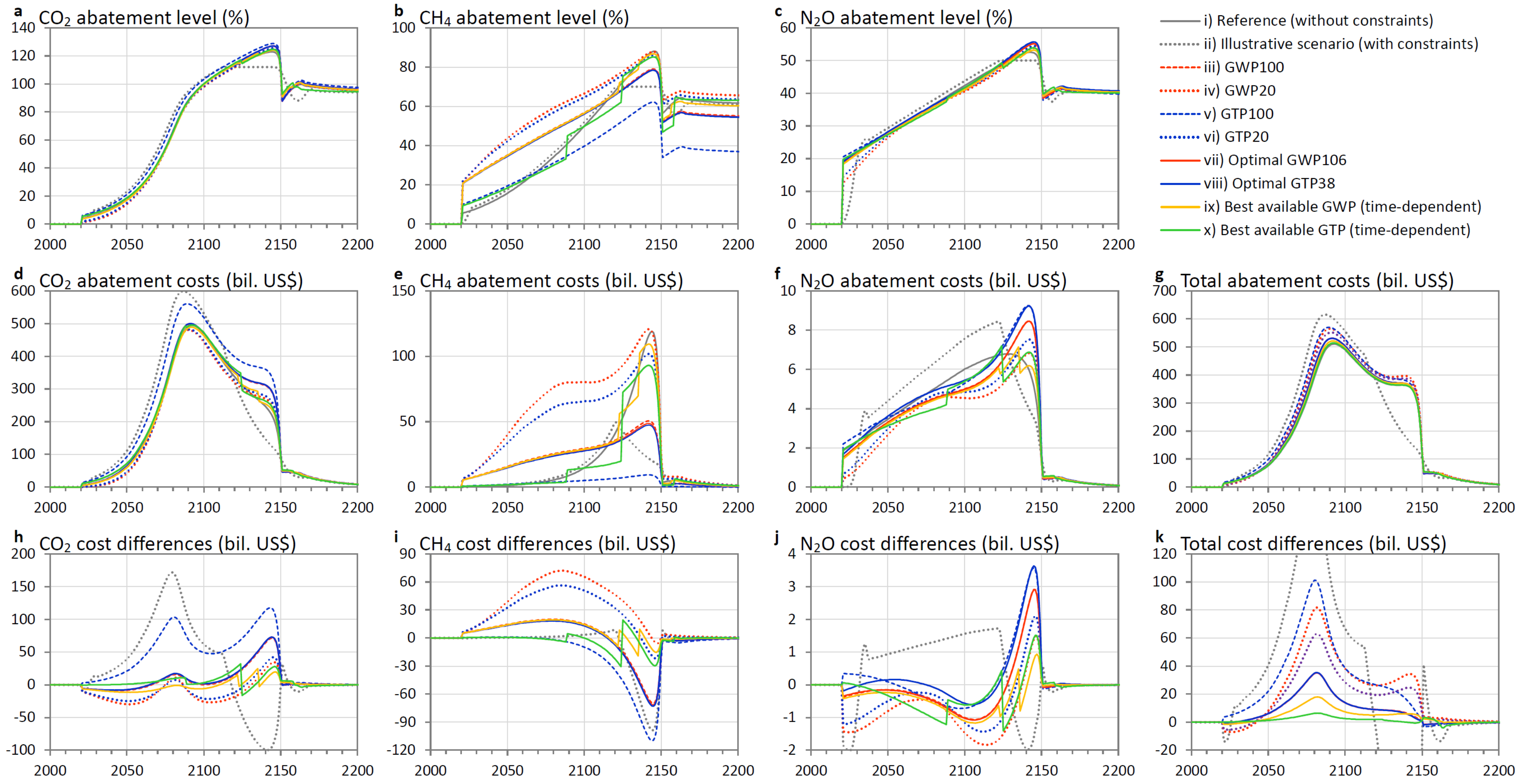

**Supplementary Fig. 14. Complete reporting of the outcomes for metric cost calculations under the 2 °C large overshoot pathway.** All panels are presented in the same way with those in Supplementary Fig. 12.

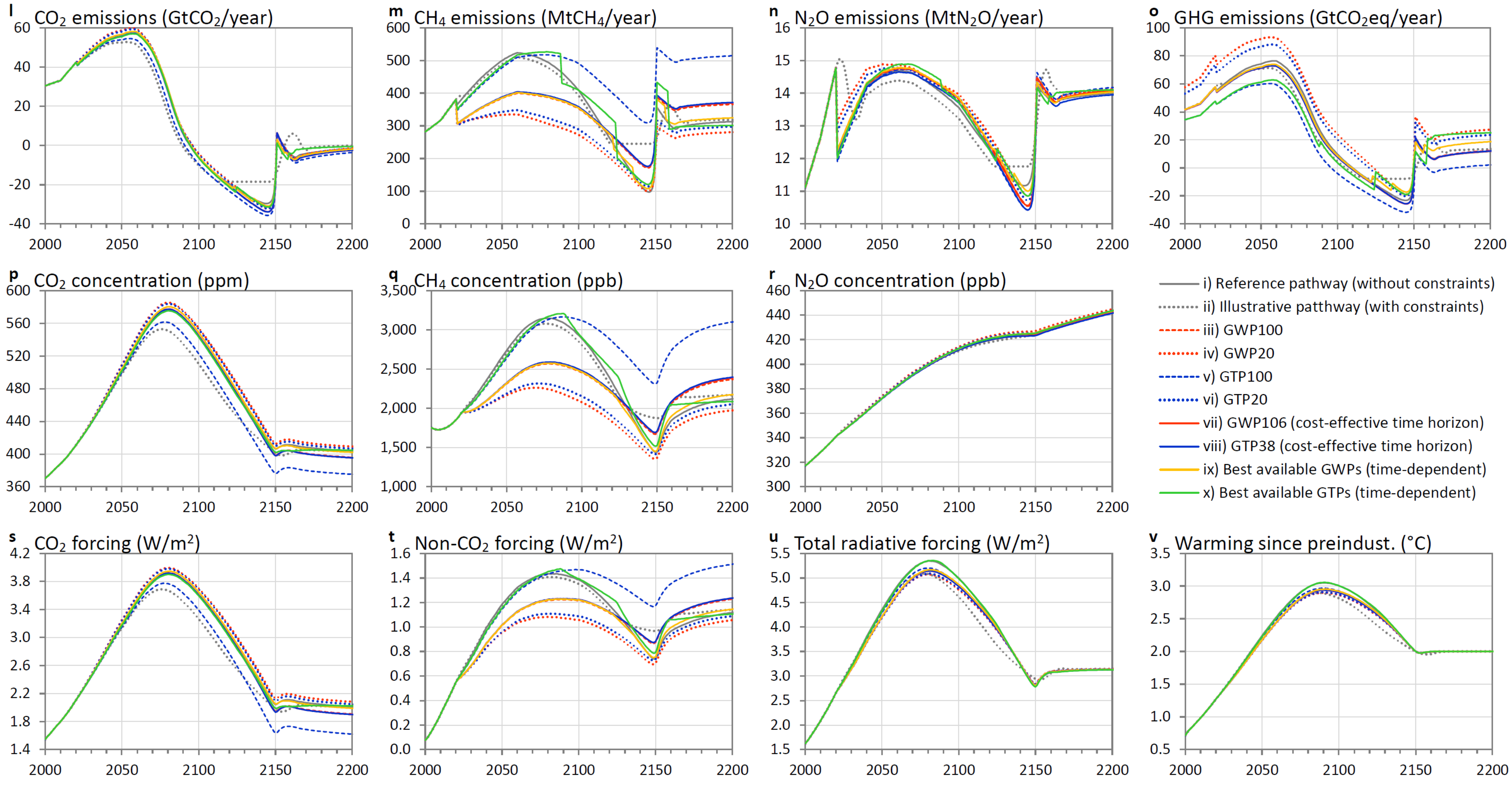

**Supplementary Fig. 14 (Cont.) | Complete reporting of the outcomes for metric cost calculations under the 2 °C large overshoot pathway.** All panels are presented in the same way with those in Supplementary Fig. 12.

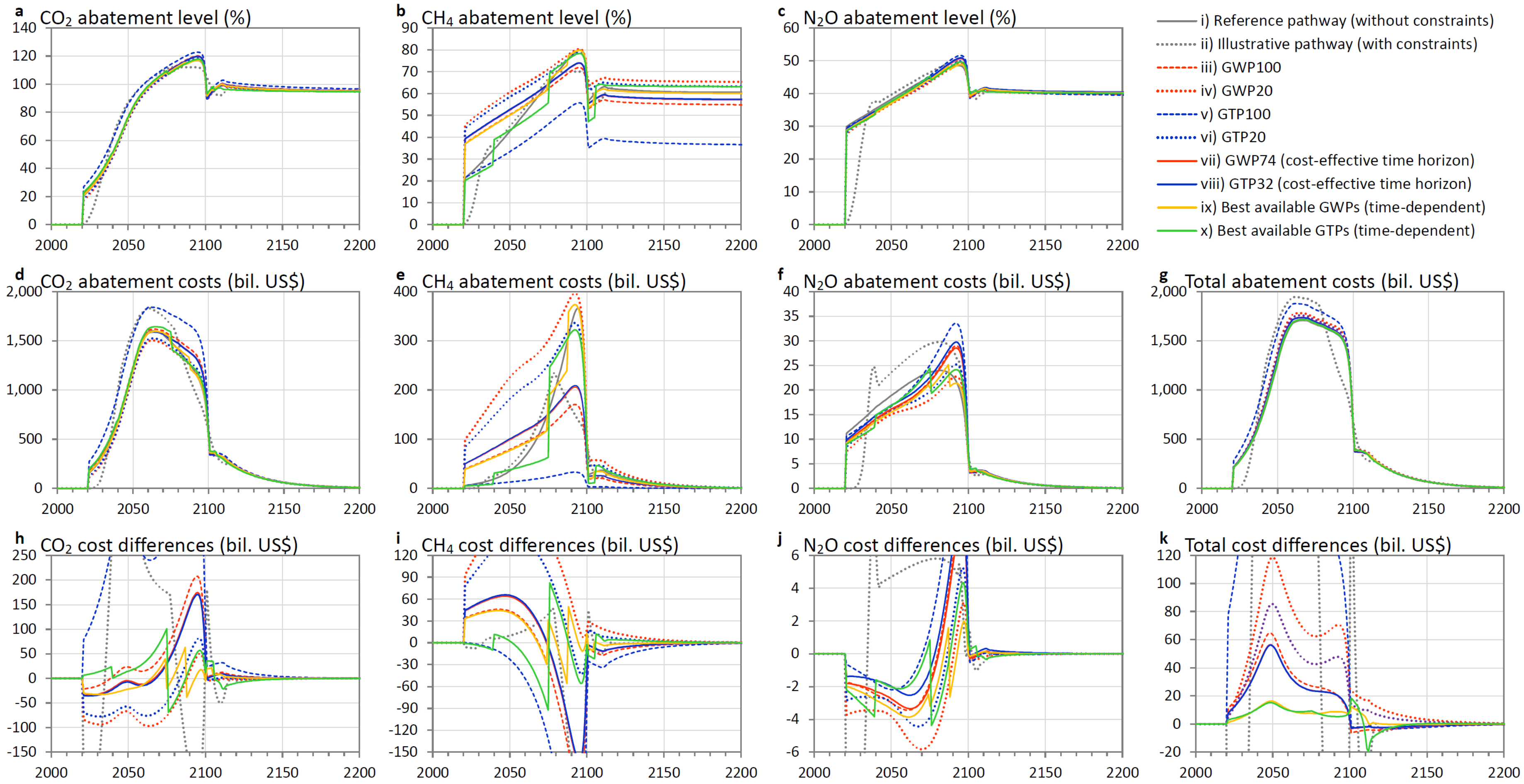

**Supplementary Fig. 15. Complete reporting of the outcomes for metric cost calculations under the 1.5 °C medium overshoot pathway.** All panels are presented in the same way with those in Supplementary Fig. 12.

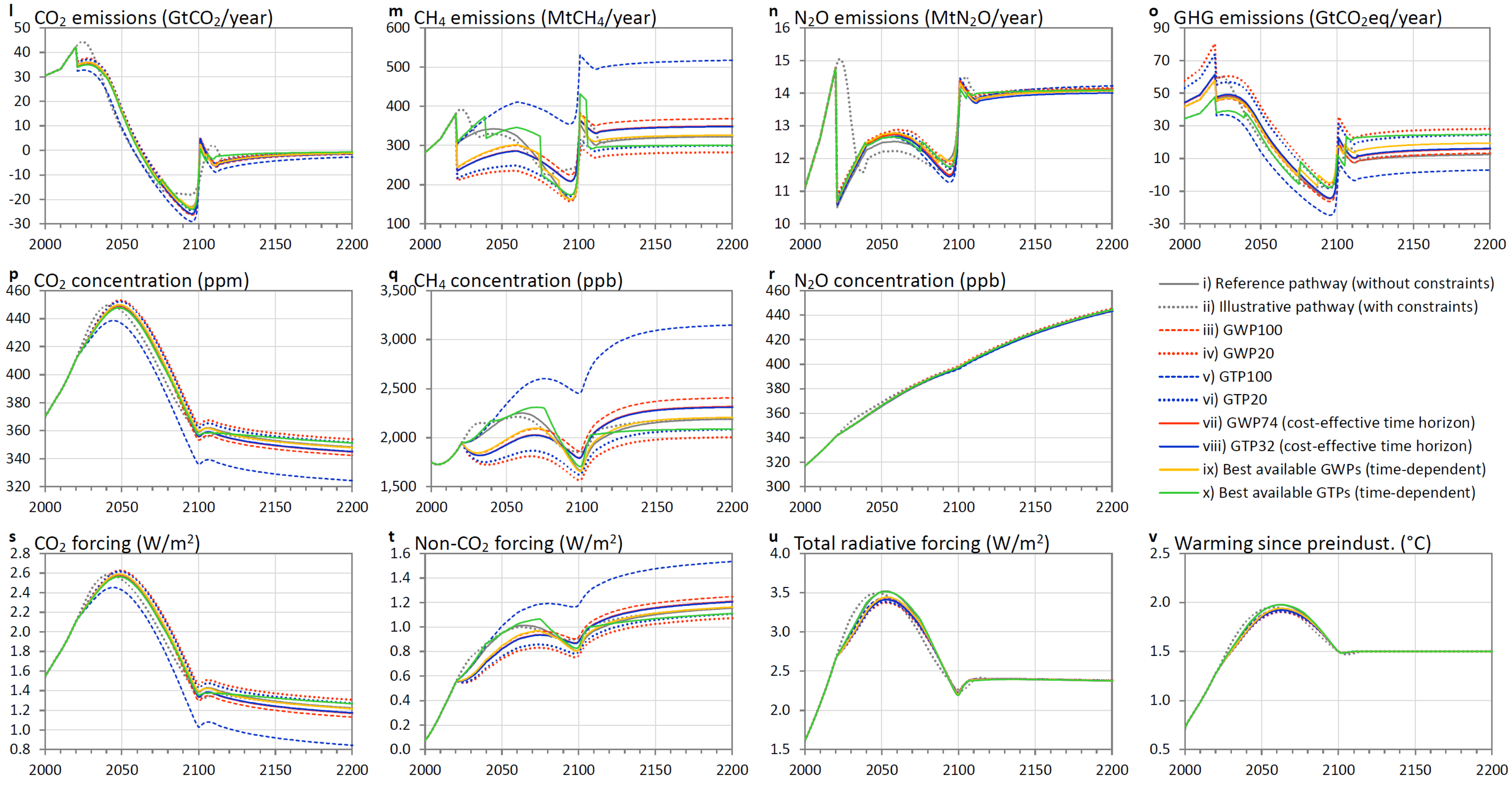

**Supplementary Fig. 15 (Cont.) | Complete reporting of the outcomes for metric cost calculations under the 1.5 °C medium overshoot pathway.** All panels are presented in the same way with those in Supplementary Fig. 12.

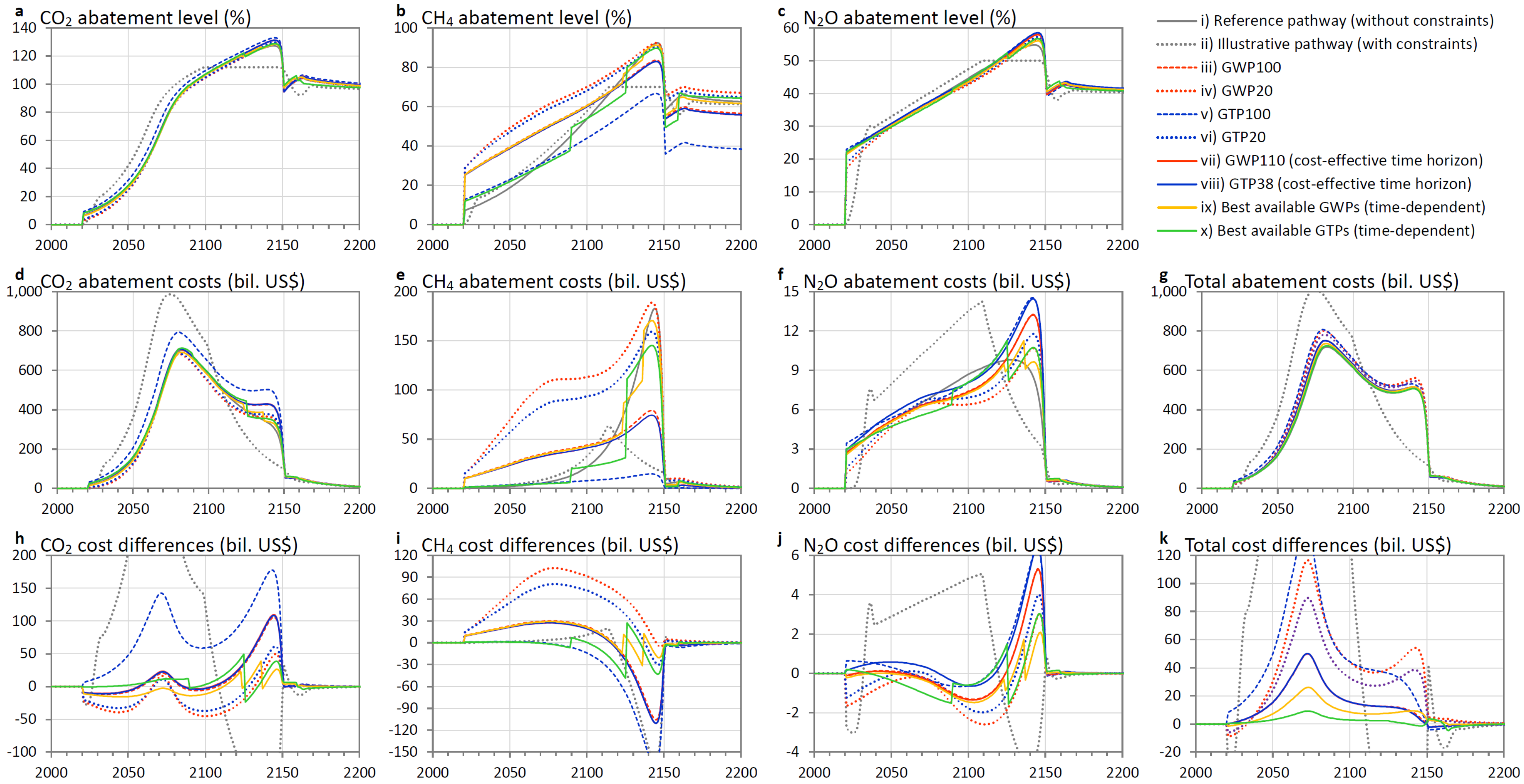

**Supplementary Fig. 16. Complete reporting of the outcomes for metric cost calculations under the 1.5 °C large overshoot pathway.** All panels are presented in the same way with those in Supplementary Fig. 12.

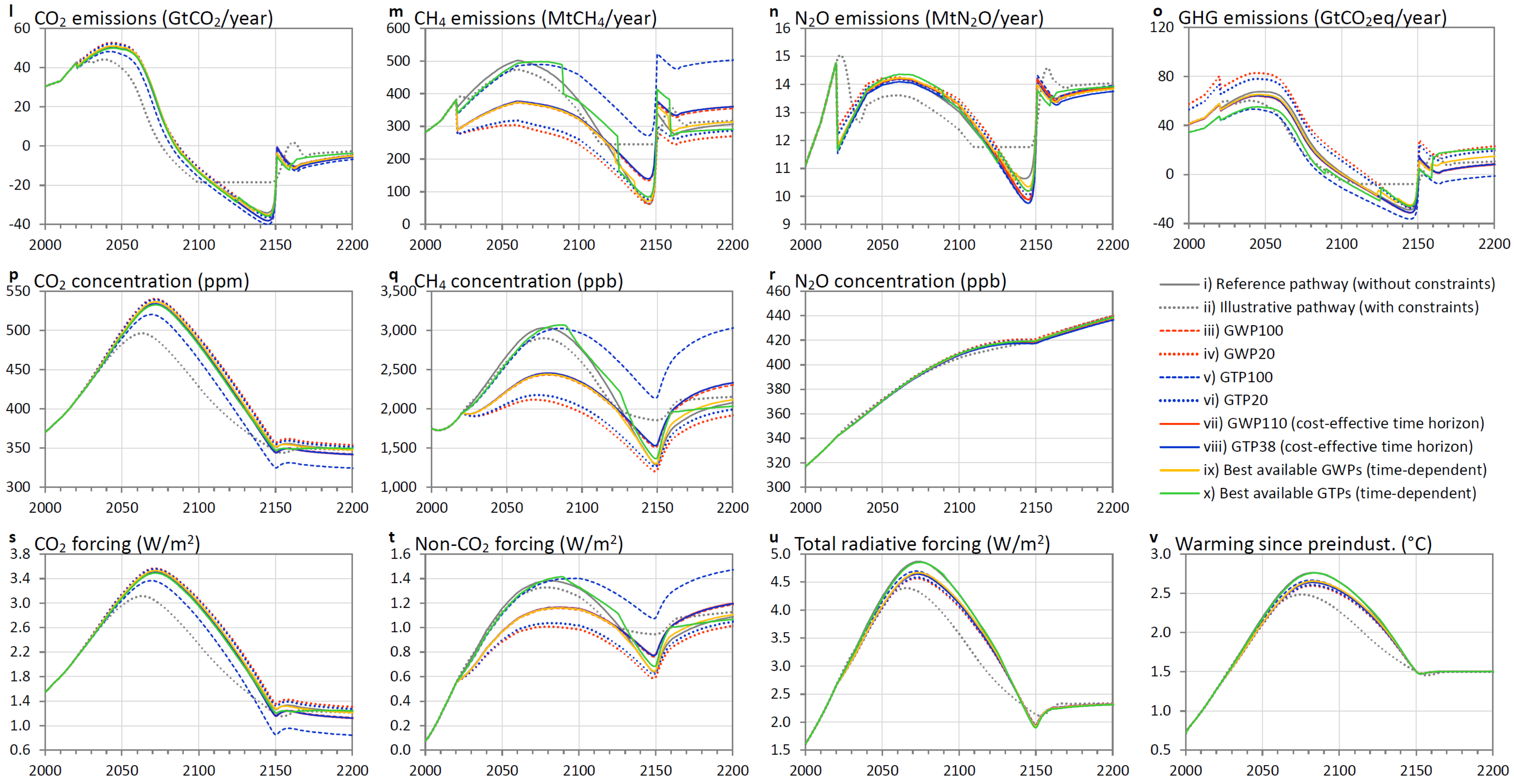

**Supplementary Fig. 16 (Cont.) | Complete reporting of the outcomes for metric cost calculations under the 1.5 °C large overshoot pathway.** All panels are presented in the same way with those in Supplementary Fig. 12.

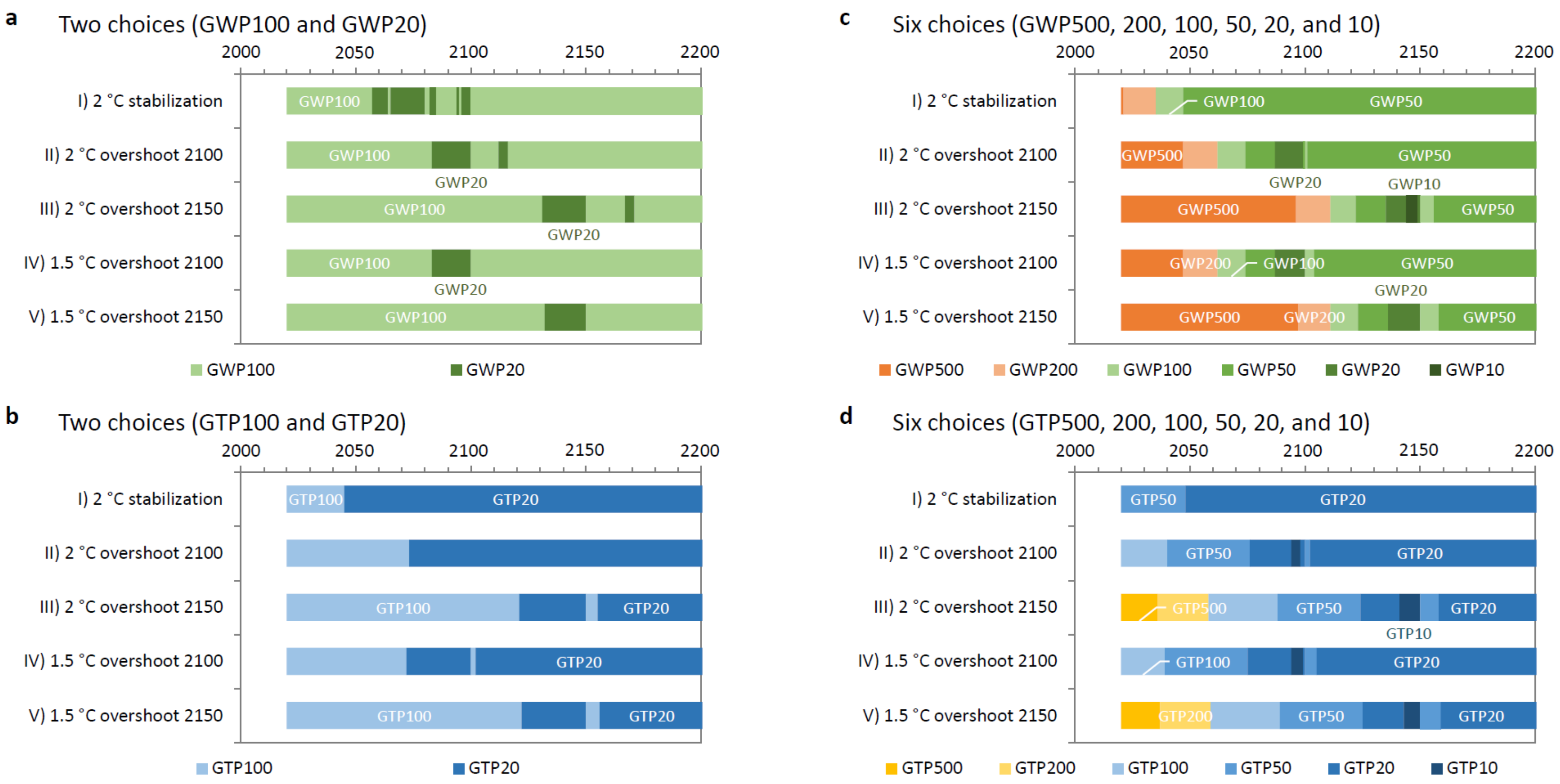

**Supplementary Fig. 17. Sensitivity analysis of the best available GWPs and GTPs with respect to the choice of available metrics.** This figure shows the sensitivity analysis of the results presented in Fig. 4 of the main paper with respect to the choice of available metrics. In the default case, three time horizons are available for GWP and GTP: 100, 50, and 20 years. See Methods for further details in the sensitivity analysis.

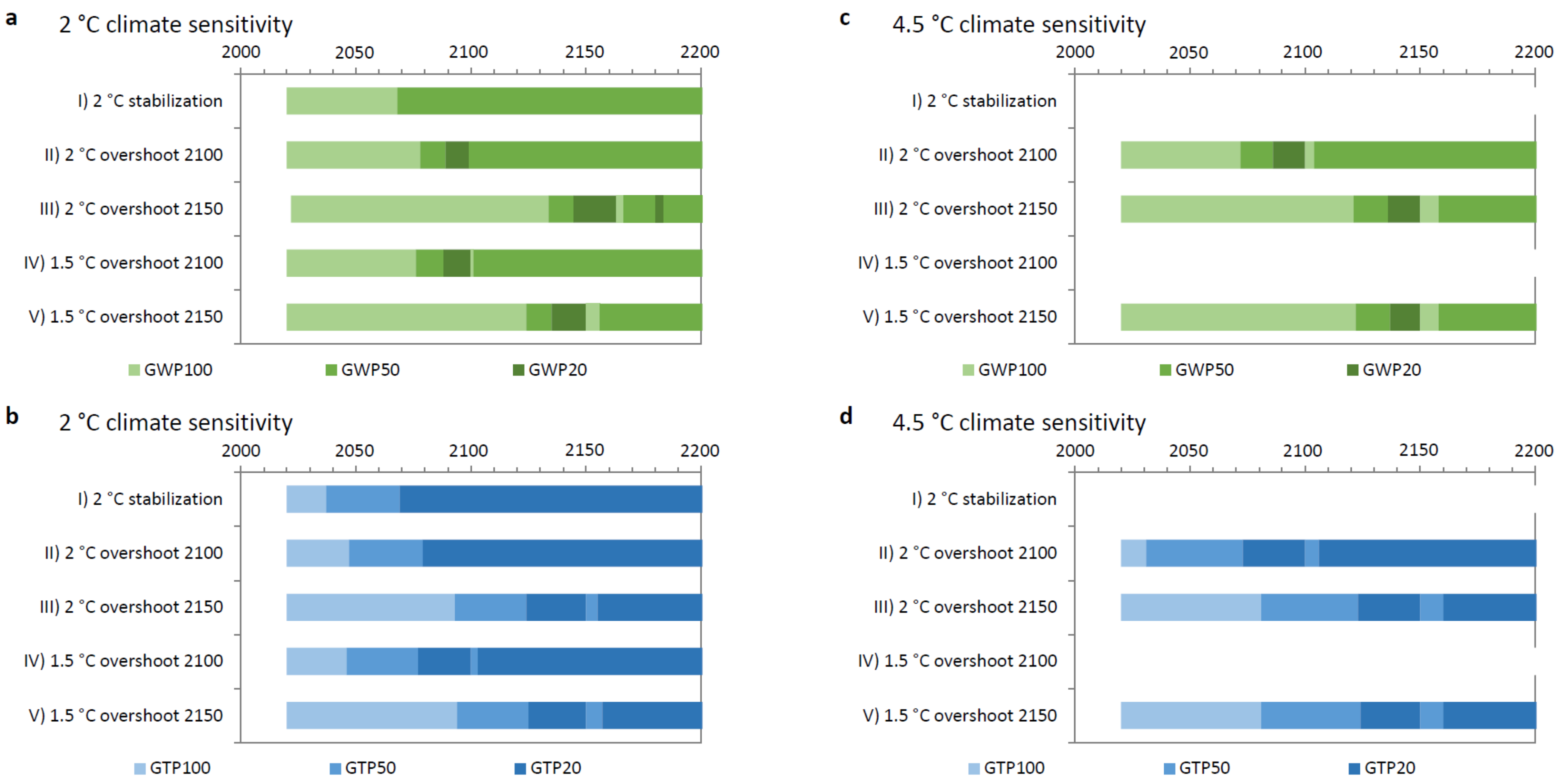

**Supplementary Fig. 18. Sensitivity analysis of the best available GWPs and GTPs with respect to the equilibrium climate sensitivity.** This figure shows the sensitivity analysis of the results presented in Fig. 4 of the main paper with respect to the equilibrium climate sensitivity. In the default case, a climate sensitivity of 3 °C is assumed. See Methods for further details in the sensitivity analysis.

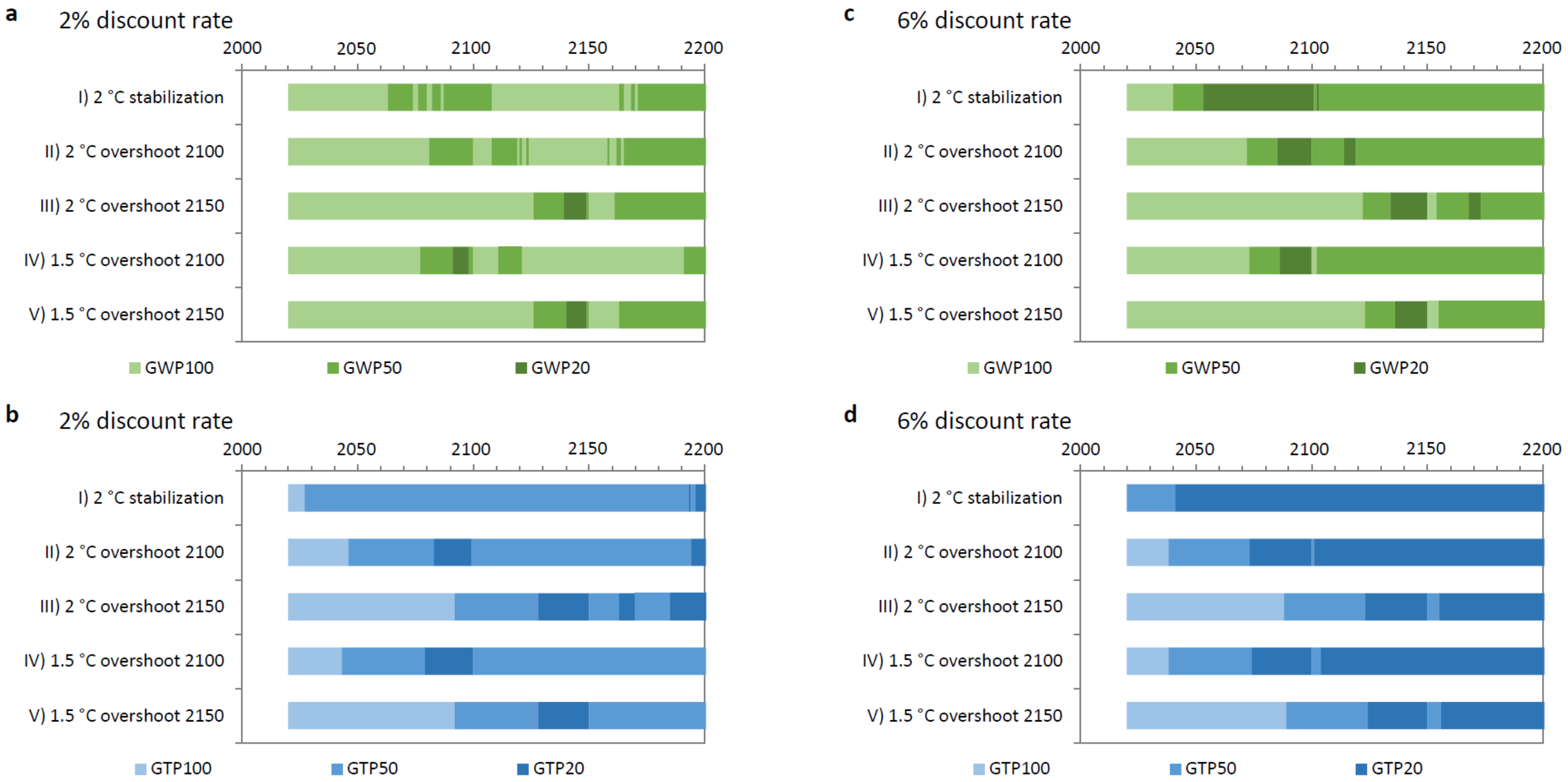

**Supplementary Fig. 19. Sensitivity analysis of the best available GWPs and GTPs with respect to the discount rate.** This figure shows the sensitivity analysis of the results presented in Fig. 4 of the main paper with respect to the discount rate. In the default case, a discount rate of 4% is assumed. See Methods for further details in the sensitivity analysis.

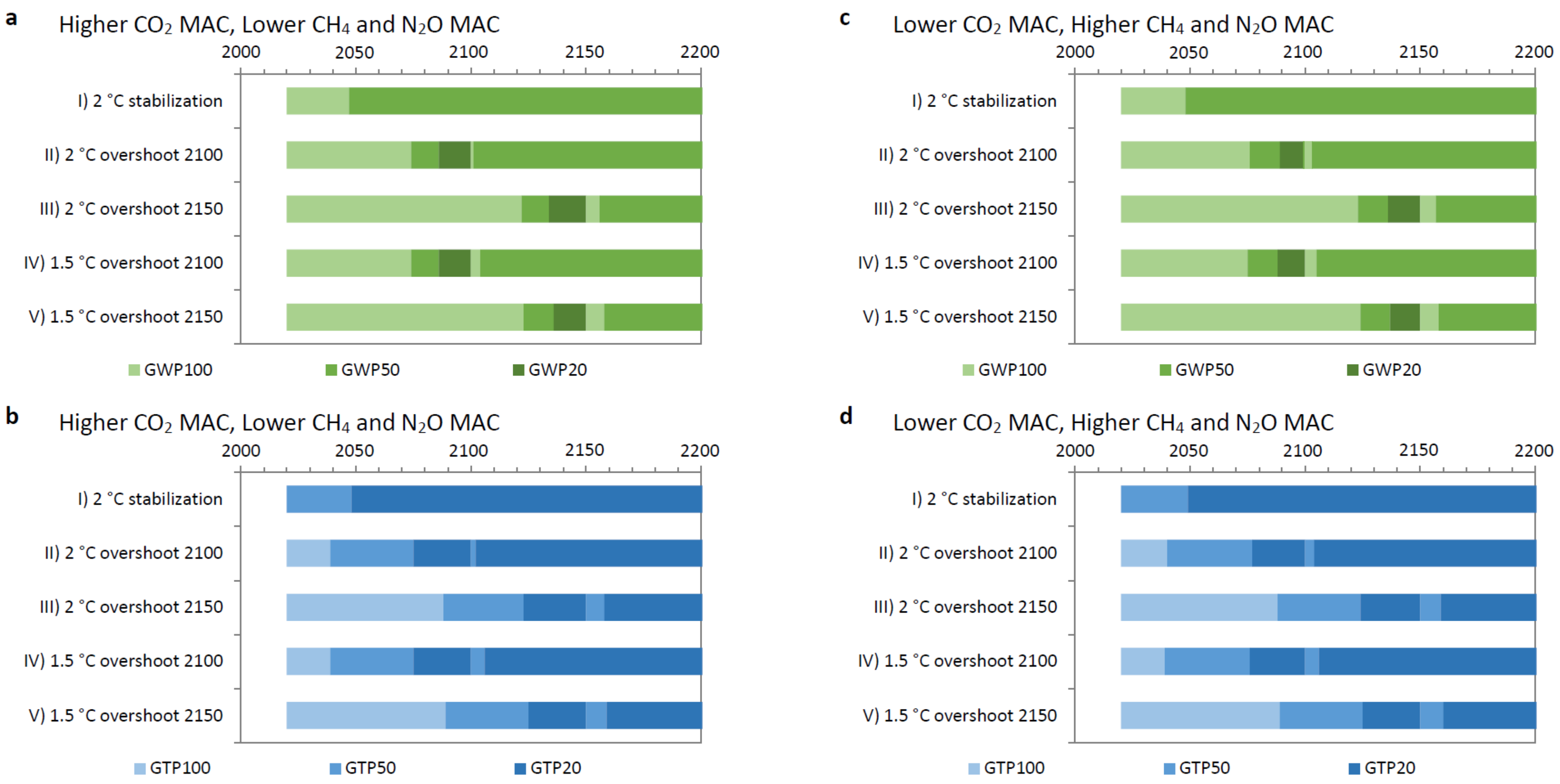

**Supplementary Fig. 20. Sensitivity analysis of the best available GWPs and GTPs with respect to the MAC curves.** This figure shows the sensitivity analysis of the results presented in Fig. 4 of the main paper with respect to the MAC curves. On the left panels, the $CO_2$ MAC curve is assumed to be 50% higher than the default curve (Supplementary Fig. 2). The $CH_4$ and $N_2O$ curves are, on the other hand, 50% lower than the default curves. On the right panels, the opposite assumptions are made for these MAC curves. See Methods for further details in the sensitivity analysis.